\documentclass[preprint,12pt,authoryear]{elsarticle}

\usepackage{amssymb}
\usepackage{amsmath}

\usepackage{url}

\usepackage{amsmath}
\usepackage{subfigure}
\usepackage{booktabs}
\usepackage{graphicx}
\usepackage{grffile}
\usepackage{amsmath,amsfonts,amssymb,amsthm,amssymb}

\usepackage{stackengine}

\newbox\subfigbox


\usepackage{listings}
\usepackage[font=small]{caption}

\theoremstyle{plain}

\theoremstyle{definition}

\theoremstyle{remark}

\usepackage{multirow}
\usepackage[table,xcdraw]{xcolor}
\usepackage{amsmath}
\usepackage{booktabs}
\usepackage{amsmath,amsfonts,amssymb,amsthm,amssymb}
\usepackage{graphicx}
\usepackage{subcaption}
\usepackage{xcolor}
\usepackage{multirow}
\usepackage{tikz}
\usetikzlibrary{decorations.pathmorphing}
\usepackage{pgf}
\usepackage{algorithmic}
\usepackage{verbatim}

\usepackage{comment}
\usepackage[all,cmtip]{xy}
\usepackage{tikz-cd}
\usetikzlibrary{shapes,arrows}
\usetikzlibrary{automata,positioning,arrows}

\usepackage{apalike}

\usepackage{setspace}

\usepackage{epstopdf}
\usepackage{subfigure}

\usepackage{stackengine}

\journal{Ocean Modelling}

\begin{document}





\begin{frontmatter}



\title{Current effects on wind generated waves near an Ocean Eddy Dipole}




\author[inst1]{Nelson Violante-Carvalho}
\author[inst1,inst2]{Thiago de Paula}
\author[inst3]{Leandro Calado}
\author[inst4]{Felipe Marques dos Santos}
\author[inst5]{Luiz Mariano Carvalho}
\author[inst6]{Andre Luiz Cordeiro dos Santos}
\author[inst7]{Wilton Z. Arruda}
\author[inst8]{Leandro Farina}

\affiliation[inst1]{organization={Ocean Engineering Program, Rio de Janeiro Federal University (COPPE-UFRJ), Brazil}}          
\affiliation[inst2]{organization={Department of Oceanic Engineering Technology, Petrobras Research and Development Center (CENPES), Brazil}}
\affiliation[inst3]{organization={Admiral Paulo Moreira Marine Studies Institute, Brazilian Navy (IEAPM), Brazil}}
\affiliation[inst4]{organization={Marine Autonomous and Robotic Systems, National Oceanography Centre Southampton (NOCS), UK}}
\affiliation[inst5]{organization={Mathematics Institute, Rio de Janeiro State University (UERJ), Brazil}}
\affiliation[inst6]{organization={Mathematics Institute, Federal Center for Technological Education of Rio de Janeiro (CEFET-RJ), Brazil}}
\affiliation[inst7]{organization={Mathematics Institute, Rio de Janeiro Federal University (UFRJ), Brazil}}
\affiliation[inst8]{organization={Institute of Mathematics and Statistics, Federal University of Rio Grande do Sul (UFRGS), Brazil}}

\begin{abstract}
Ocean eddy dipoles are among the most common mesoscale features and may be ubiquitous across the global oceans.
However, wave-current interactions in their proximity have not been extensively studied.
Here we examine the impact of surface currents on the wave field near an ocean eddy dipole. Using the WW3 wave model, we conducted idealized numerical simulations to assess the influence of different configurations on the spatial variability of Significant Wave Height ($H_s$). Additionally, a two-month hindcast of a strong dipole event in the Southwestern Atlantic Ocean was performed using three distinct surface current products: SSalto/Duacs, HYCOM NCODA and GlobCurrent. Among these, HYCOM, which incorporates ageostrophic effects, provided a more detailed representation of oceanic energy compared to GlobCurrent and SSalto/Duacs, which primarily reflect geostrophic components. 
The hindcast assessment employed denoised altimeter-derived $H_s$ data, with a spatial resolution of approximately 6~km.
The greatest increase in wave energy occurs in the region between the peak values of positive and negative vorticity, where the opposing surface currents reach their maximum intensity.
Therefore, dipoles act as converging lenses for surface waves, channeling their refraction towards the central jet.
Despite its poorer spatial and temporal resolutions, SSalto-Duacs surface current data provides more reliable $H_s$ fields, in the study region where geostrophic dynamics are expected to be significant or even dominant. 
Both Absolute Dynamic Topography derived surface current inputs produce comparable effects on the wave field, with the inclusion of the Ekman component yielding no substantial enhancement.
HYCOM captures a broader range of dynamical processes, essential for accurately representing the total energy, though discrepancies with SSalto/Duacs data may arise from assimilation inaccuracies and model limitations. 
While gridded altimetry may underestimate total current components during dipole events, it offers precise insights into their positioning and evolution, useful for specific event analysis and near real-time forecasting for marine safety.
\end{abstract}

\end{frontmatter}

\section{Introduction}\label{sec_intro}

Wind generated waves are affected by underlying currents and this interaction becomes particularly complex in the vicinity of ocean eddy dipoles. 
These mesoscale features, characterized by rotating currents with opposing polarities, can significantly influence wave propagation and dynamics, including changes in wave period, direction and energy distribution.
Understanding how currents modify waves within eddy dipoles is crucial for ocean engineering, ship routing and hazard assessments. 
Additionally, studying current effects on waves can provide insights into broader air-sea momentum exchange processes and their role in ocean circulation patterns \citep{BENETAZZO2013152, Marechal-Ardhuin-2021, https://doi.org/10.1029/2024JC021581}.

Wave–current interactions have been investigated in recent studies, including, among others, \cite{fab2, fab, QUILFEN2018561, romero2, ROMERO2020101662, bia, lentini}. 
The change of wave height induced by surface currents depends on their relative velocities~---~if the group velocity of the waves is much larger than the current speed, the waves will propagate unaffectedly.
However, if the surface current speed is a considerable fraction of the group velocity, the waves will be advected and hence their propagation will be affected by the current.
\cite{onorato} demonstrated with numerical simulations, employing a modified nonlinear Schrodinger equation, that the maximum amplitude of monochromatic waves in opposing currents can increase significantly when the ratio between surface current speed and group velocity is in the range 0.1 to 0.4.
For relative larger currents, and therefore larger ratios, wave breaking and blocking might occur.
Currents can modify the direction of wave propagation at a given location through the combined effects of refraction and wave action advection. The degree of deviation from great circle paths is not directly dependent on the surface current speed but rather on the ratio between the vertical component of vorticity and the group velocity.

The primary dynamic component of surface circulation in the South Brazilian Bight is the Brazil Current (BC), a western boundary current that completes the wind-driven South Atlantic Subtropical Gyre. 
As it flows south-southwestward along the continental slope, the BC exhibits a pronounced meandering pattern that frequently encloses eddies. 
These eddies are primarily generated by flow-topography interactions and baroclinic instability \citep{DASILVEIRA2008187, https://doi.org/10.1002/2013JC009143, silveira2023brazil}. 
Recurrent eddy-like features have been documented along the entire path of the BC, from its genesis near 15°S to the Brazil–Malvinas Confluence at 39°S \citep{Bilo2014}. 
The most persistent eddies are typically observed off Vitoria (20°S), Cabo de São Tome (22°S), and Cabo Frio (23°S) \citep{calado2006parametric}. 
In the region where the Cabo Frio Eddy forms (around 23°S), the BC encounters a significant change in coastline orientation, shifting from a southwest to a westward direction over a span of a few tens of kilometers. 
Due to its inertia, the BC veers away from the shelf break and is displaced to greater depths. 
This displacement leads to the acquisition of cyclonic relative vorticity as the current conserves its potential vorticity \citep{Campos2000}. 
The resulting cyclonic circulation at this location occasionally couples with westward-propagating anticyclonic features, resulting in the formation of an eddy dipole structure. 
This structure is characterized by a strong intensification of surface currents in its central region \citep{thiago}. 
The dipoles formed in this area commonly propagate south-southwestward, substantially altering local surface circulation patterns.

\begin{figure}[t!]
\centering
\includegraphics[width=1.0\textwidth]{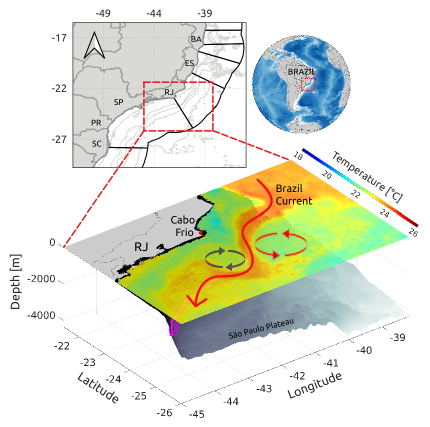}
\caption{\small{Key topographic features and main surface circulation with an example of a typical dipole composed of a cyclonic eddy (black) and an anticiclonic eddy (red).}}
\label{fig:fig1aa}    
\end{figure}

One of the defining characteristics of ocean eddy dipoles is the formation of a central jet, characterized by its high velocity and narrow width, see a schematic representation over the South Brazilian Bight in Figure~\ref{fig:fig1aa}.
When the eddies are in close proximity, their rotational flows converge, creating a narrow, intense current between them. 
The central jet typically exhibits higher velocities compared to the surrounding eddy flows~---~the speed of the jet can reach up to several tens of centimeters per second, depending on the intensity and size of the eddies.
\cite{Ni2020} pointed out that dipoles are widespread in the global ocean, with about 30–40\% of the mesoscale eddies identified in altimeter data as dipoles~---~the percentage is even higher in energetic regions such as the Gulf Stream and the Southern Ocean. 

Dipoles create complex spatial variations in the surface current field, leading to wave refraction. 
As waves propagate through the variable flow fields of the cyclonic and anticyclonic eddies, their trajectories bend according to the changes in current velocity and direction. 
This refraction can lead to wave focusing or defocusing, depending on the relative orientation of the waves and the dipole. 
The wave focusing effect, particularly along the central jet, can result in increased wave heights and energy concentration.

We examine the response of the spatial variability of the wave field to the surface currents in the vicinity of an eddy dipole employing remote sensing data and numerical models. %
As a first approach, we analyze an idealized setup employing the numerical wave model WAVEWATCH III \citep[WW3, see][]{ww3} forced by realistic surface currents to discuss how the significant wave height ($H_s$) field is affected.
Additionally, a two-month hindcast of a very intense dipole event is examined, with three distinct surface current fields used as input to the WW3 wave model 
in the Southwestern Atlantic Ocean, a region known for the recurrent occurrence of dipoles.
The Hybrid Coordinate Ocean Model (HYCOM) relies on simulations driven by physical equations, whereas GlobCurrent and SSalto integrates observational data from satellites. 
The latter two are derived from altimeter data and therefore the surface fields include only the geostrophic component~---~plus the Ekman component added in the GlobCurrent product.
Numerical models, such as HYCOM, are also essential for incorporating ageostrophic effects not present in the altimeter data. 
Each of the three surface current products differs in terms of methodologies, data sources, (spatial and temporal) resolutions and applications. 
We discuss the impacts of using these distinct products as input to the hindcasts of the wave field. 

\section{The Effects of Currents on Waves in the Vicinity of Eddy Dipoles}\label{back}
\cite{PEREGRINE19769} highlighted the occurrence of frequency Doppler shift when waves propagate over currents, discussing the interplay between the intrinsic frequency of the wave and the relative motion induced by the current, in the form
\begin{equation}\label{sec:mmodel:eq:doppler}
	\omega=\sigma + \mathbf{k}\cdot\mathbf{U}.%
\end{equation}
The absolute frequency as observed from a fixed reference frame is $\omega$, $\sigma$ is the intrinsic frequency when observed from a moving reference frame, $\mathbf{k}$ is the wavenumber vector and $\mathbf{U} = \mathbf{U}(t, \mathbf{x})$ represents the horizontal surface current, with $t$ denoting time and $\mathbf{x}$ representing the position vector.

The temporal changes in the current field occur at a rate much slower than that of propagating waves. 
Therefore, the temporal scales of the current field are significantly longer than the corresponding wave scales. 
The assumption extends to the spatial domain, with scales associated with current variations significantly larger than that of the waves, hence 
\begin{equation}\label{sec:mmodel:eq:hypo}
	k\gg\max\Bigg|\frac{1}{U}\frac{\partial U}{\partial x}\Bigg|\;\text{ and }\;\omega\gg\max\Bigg|\frac{1}{U}\frac{\partial U}{\partial t}\Bigg|.%
\end{equation}


Equation (\ref{sec:mmodel:eq:doppler}) underscores the dynamic nature of wave propagation over currents. 
The term $\mathbf{k} \cdot \mathbf{U}$ denotes the influence of the current velocity on the wave frequency and could be perceived as a Doppler shift; when current and wave propagate in the same direction ($\mathbf{k} \cdot \mathbf{U} > 0$), the waves are stretched~---~conversely, if the current opposes the wave direction, the waves become shorter. 
When the waves propagate orthogonally to the current, no influence is observed.

The relation between the intrinsic frequency and the
wavenumber of a uniform plane wave of infinitesimal amplitude, propagating over
still water of uniform depth $h$ relies upon the dispersion relationship
\begin{equation}\label{sec:mmodel:eq:dispersion}
	\sigma^2 = gk\tanh kh,%
\end{equation}
where $g$ is the gravitational acceleration and $k=||\mathbf{k}||_2$. 
On the surface of a uniform current, the dispersion relation \eqref{sec:mmodel:eq:dispersion}, using 
Equation \eqref{sec:mmodel:eq:doppler}, turns into
\begin{equation}\label{sec:mmodel:eq:dispersioncurr}
	(\omega-\mathbf{k}\cdot\mathbf{U})^2 = gk\tanh kh.%
\end{equation}
Considering the particular case of waves collinear with currents, the amplitude in the presence of currents ($a$) and the amplitude without currents ($a_0$) have the ratio \citep{phillips}
\begin{equation}\label{sec:mmodel:eq:amplratio_phi}
	\frac{a}{a_0}=\frac{c_{0}}{\sqrt{(2U+c)c}},
\end{equation}
where $c$ and $c_0$ are, respectively, phase speed with or without currents. 
To illustrate, a 15~s period wave propagating over a 1~$ms^{-1}$ opposing current can result in a 10$\%$ amplification in wave height.
In the same scenario, this is enhanced to around 25$\%$ for a wave with a shorter 7~s period~---~naturally, stronger opposing currents result in stronger amplifications.
Hence, opposing waves near the dipole central jet are expected to increase in height and steepness \citep{onorato2, toffoli}.
The wave amplification increases rapidly to opposing current velocities~---~the slower the waves and the stronger the currents, the steeper the curve.

\cite{KENYON19711023} discussed the possible causes, later supported by several simulations presented by \cite{gallet2014refraction}, of the discrepancies reported by \cite{munk} in their work about swell propagation over large distances.
For instance, islands could block and diffract swell energy away from great circle propagation \citep{nvc_diff}~---~and ocean currents could potentially be accounted for by causing refraction \citep{roger}.
Currents can alter the direction of wave propagation at a specific location through both the processes of refraction and advection of wave action, not included in the analysis by \cite{munk}.

The amount of bending from great circles is not function of the surface current speed itself, but of the ratio between the vertical component of the vorticity ($\zeta$) and wave group velocity ($c_g$):
\begin{equation}\label{sec:mmodel:eq:ratioraycurv}
	\chi=\frac{\zeta}{c_g}+\mathcal{O}(e^2),
\end{equation}
where $\chi$ is the ray curvature, $\zeta = \frac{\partial v}{\partial x}-\frac{\partial u}{\partial y}$, $\mathbf{U}=(u,v)$ the current velocity and $e=||\mathbf{U}||_2/c_g$~---~see a thorough discussion in \cite{KENYON19711023} and \cite{dysthe_2001} considering predictions from ray theory.
Slower, shorter waves are, therefore, more refracted.
Moreover, the bending direction follows the signal of the vorticity field~---~positive vorticity causes waves to bend to the left and negative to the right~---~and the bending increases with increasing current vertical vorticity.
Therefore, an eddy dipole acts like a focal lens by bending waves that oppose the central jet's current, effectively redirecting them toward the jet.
The combined influence of refraction and advection over the central jet leads to a more substantial increase in wave energy compared to the effects of refraction and advection acting independently.

\section{Data}

\subsection{Surface Current Fields}

 \subsubsection{Surface Current Field Employed in the Idealized Simulations}  \label{curr_ideal}
\begin{figure}[t!]
\centering
\subfigure[]{%
\label{subfig:9a}\resizebox*{6.5cm}{!}{\includegraphics{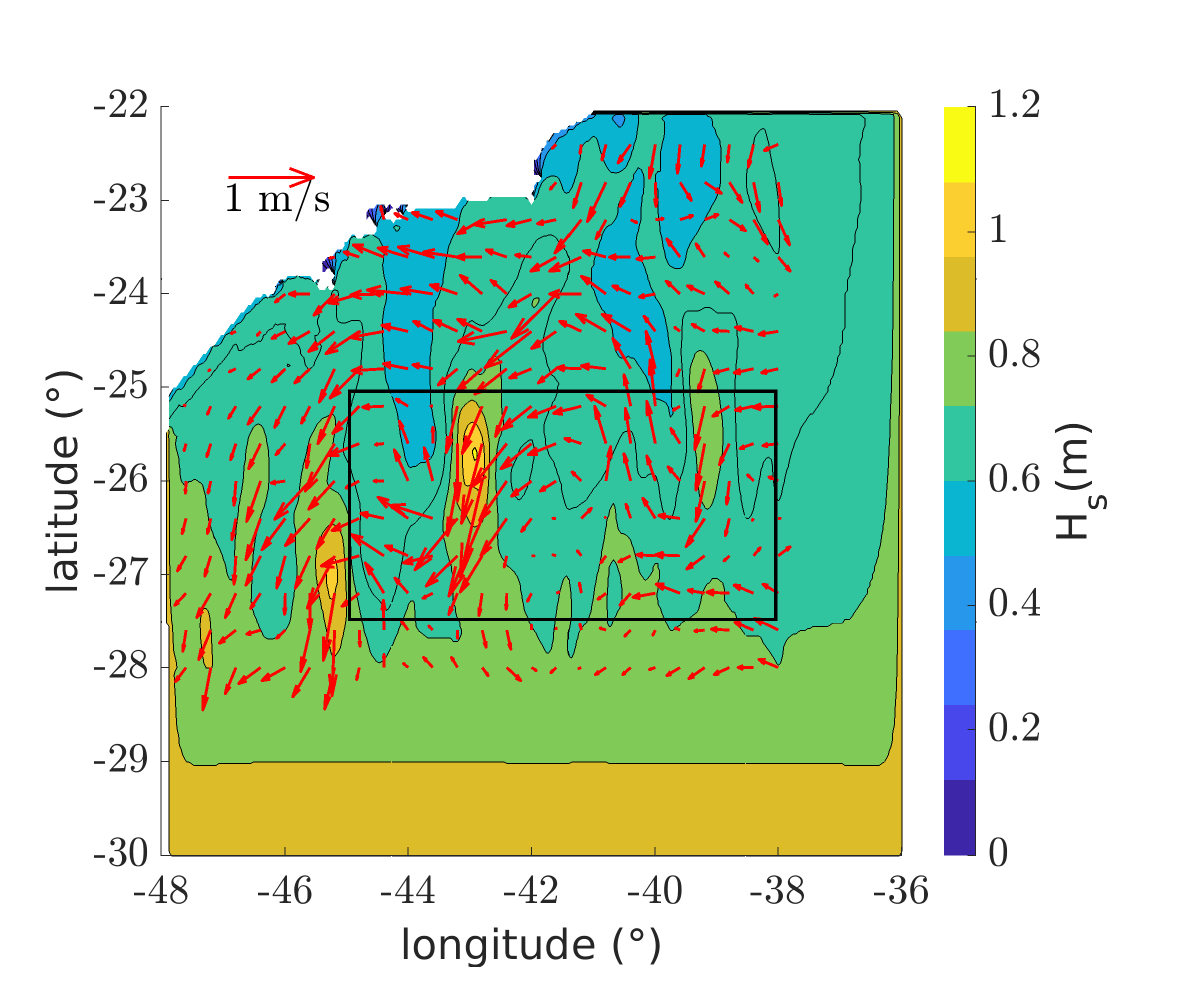}}}
\subfigure[]{%
\label{subfig:9d}\resizebox*{6.5cm}{!} {\includegraphics{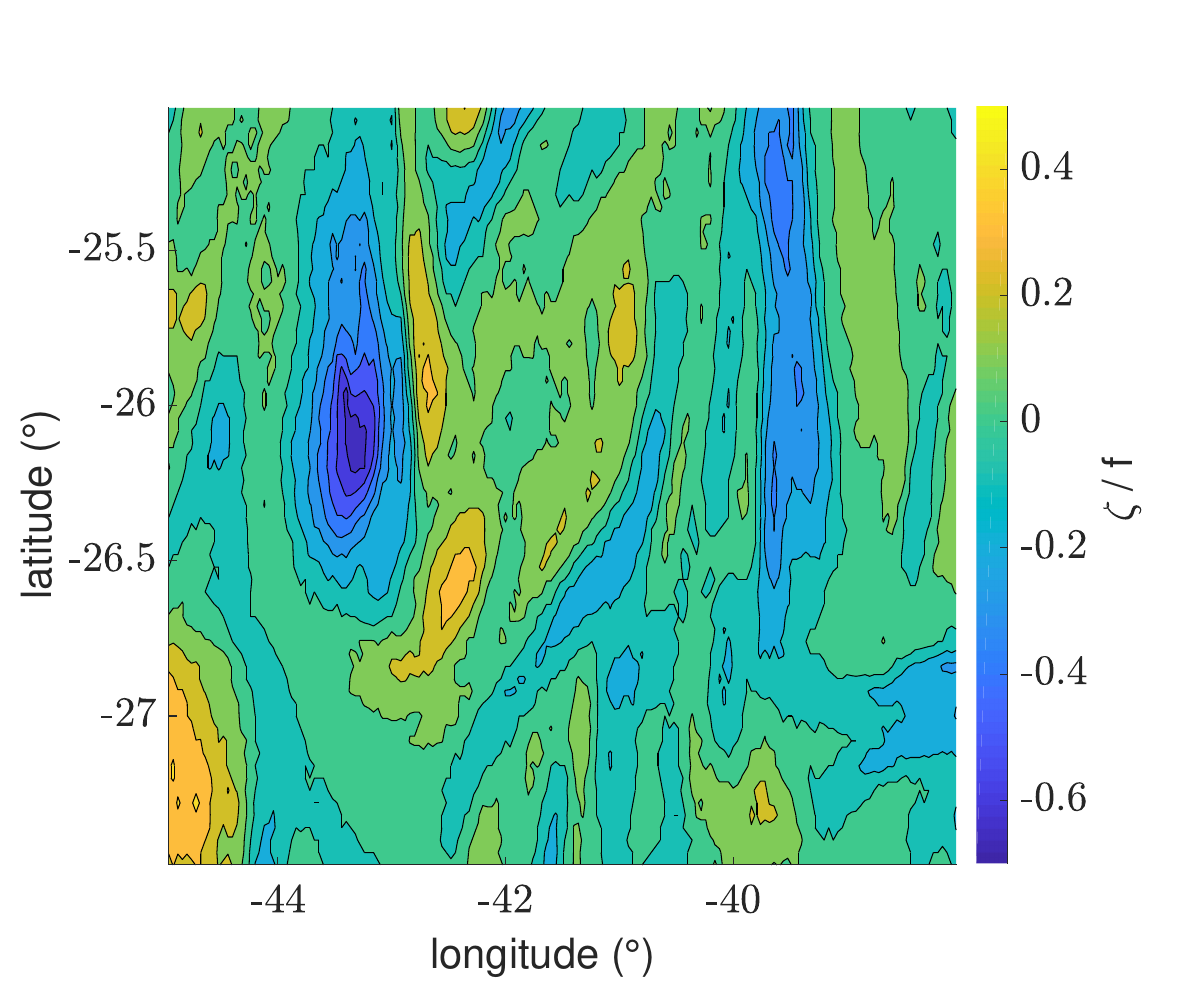}}}
\subfigure[]{%
\label{subfig:9g}\resizebox*{6.5cm}{!} {\includegraphics[width=\textwidth]{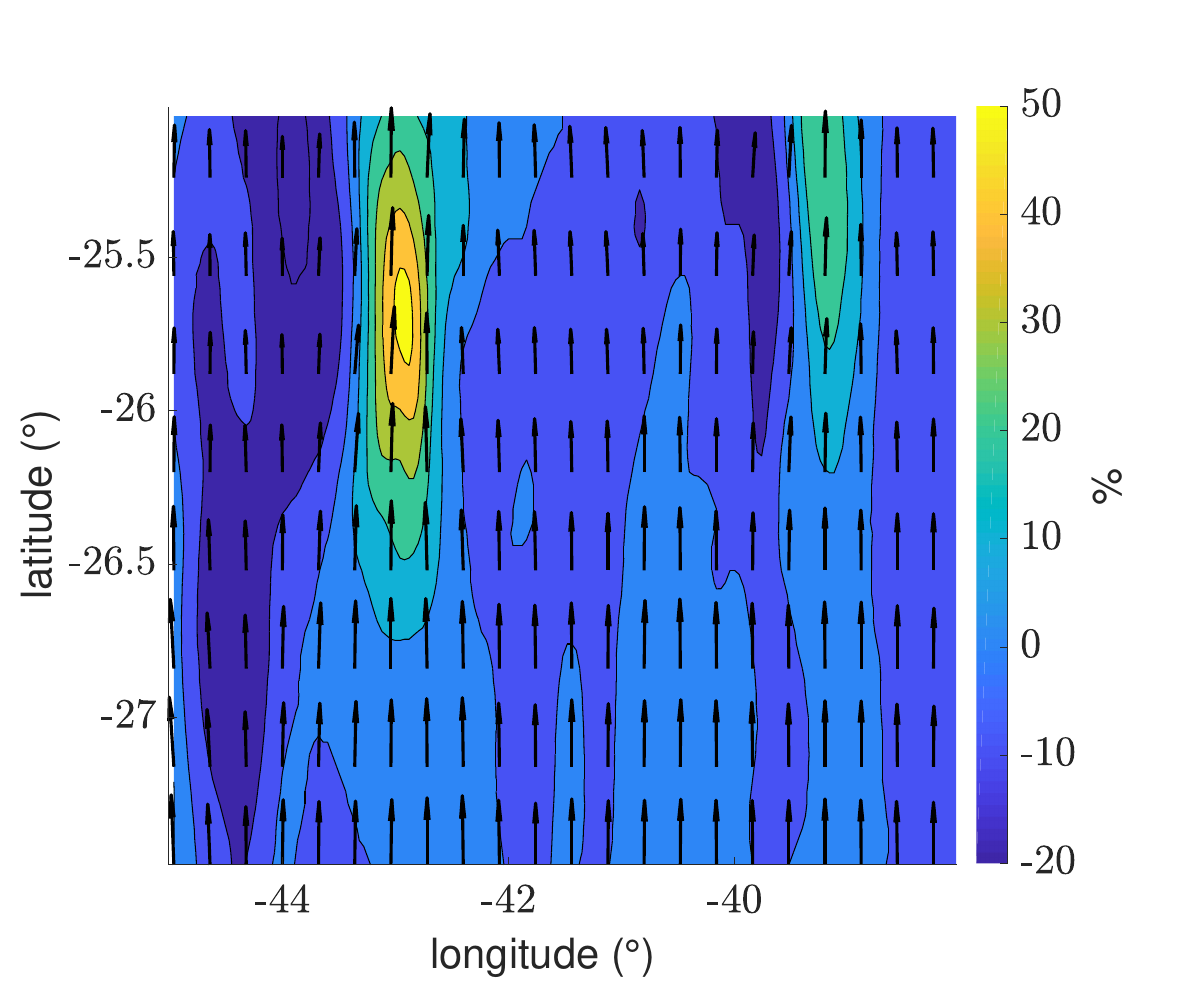}}}
\subfigure[]{%
\label{subfig:9h}\resizebox*{6.5cm}{!} {\includegraphics[width=\textwidth]{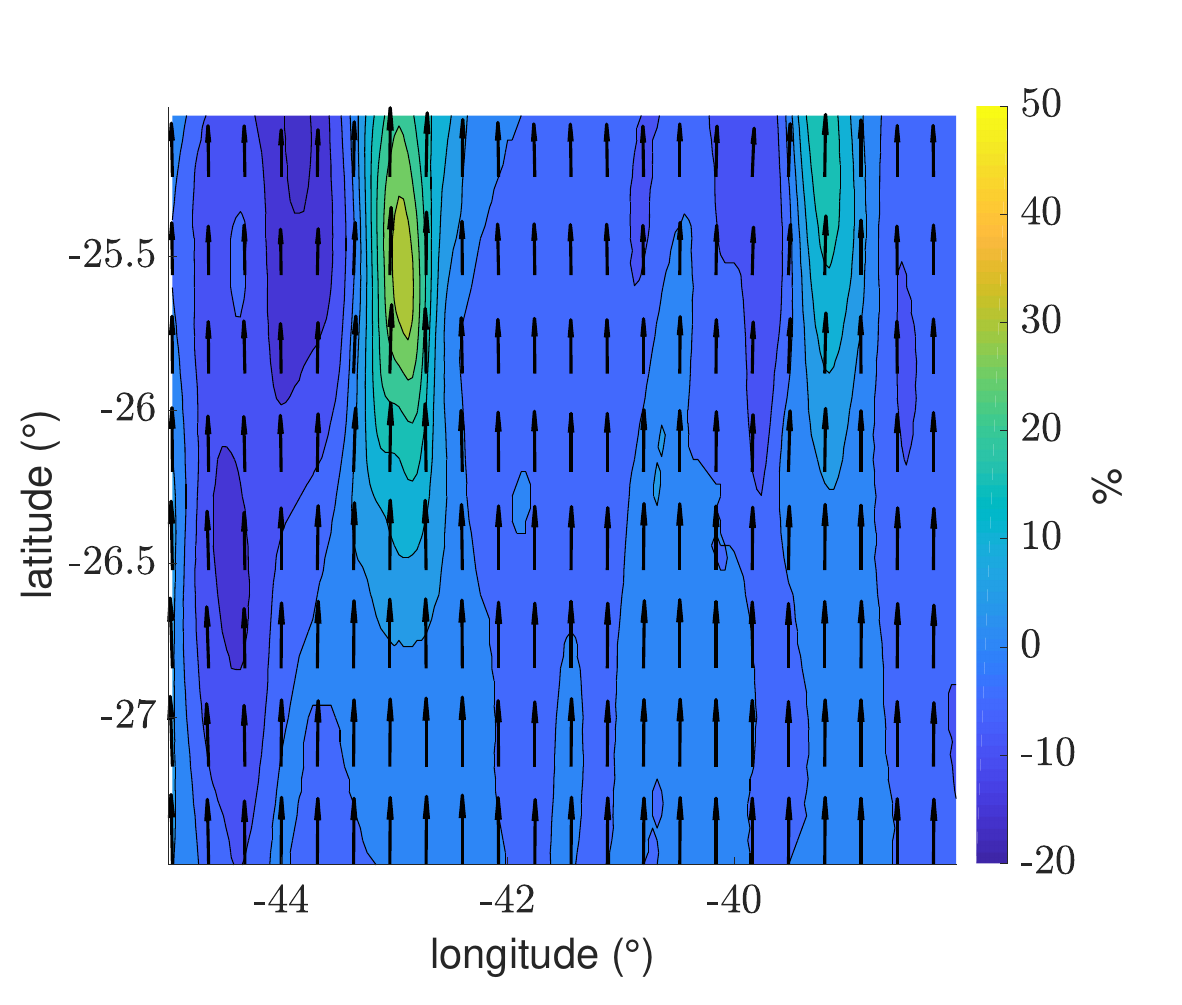}}}
\caption{\small{[a] is the map of significant wave height ($H_s$) for the 7~s waves ($\epsilon=0.0822$). The arrows indicate the surface current direction and magnitude with its scale displayed at the left top corner. [b] is the Rossby number~---~the surface relative vorticity $\zeta$ normalized by the Coriolis parameter $f$~---~in the region shown as a black rectangle in [a]. [c d] are the relative difference of $H_s$, in the black rectangle, between the runs using the current input and the run with no current (respectively for the 7~s and 15~s waves). The arrows indicate wave propagation direction scaled by $H_s$.}}
\label{fig:fig9n}
\end{figure}

A realistic surface field from a HYCOM snapshot of 08-Sep-2010 07:00:00 was employed, see Figure~\ref{fig:fig9n}a. 
The selection is based on the occurrence of an intense dipole in the Sao Paulo Plateau, where ocean currents frequently intensify. 
Furthermore, this is a particularly interesting selection for idealized simulations because the central jet presents a very well defined and straight North-South orientation.
This event is thoroughly discussed in \cite{thiago}, with speed comparisons between a regional simulation using ROMS and two global reanalyses~---~GLORYS and HYCOM~---~against mooring measurements positioned on the central axis of the dipole. 
The results demonstrated that HYCOM was capable of reproducing the magnitude of the currents, greater than 1~$ms^{-1}$. 
The velocity field was used as a series of static snapshots~---~without temporal evolution~---~to drive the wave model with horizontal resolution of 1/12$^\circ$ ($\sim$9~km) and temporal resolution of one~hour.
The steady state conditions eliminate the need to account for transient effects and time-varying behavior, therefore simplifying the analysis.


\subsubsection{Surface Current Field Employed in the Two-Month Hindcast}\label{3cur}
Three distinct surface current fields were employed as input for the wave model hindcasts, comprising the period of August and September 2010.
The main differences between SSalto/Duacs, HYCOM NCODA and GlobCurrent lie in the methods and data sources used to derive these fields~---~and consequently their spatial and temporal resolution, accuracy and typical applications.

The SSalto/Duacs from AVISO (Archiving, Validation and Interpretation of Satellite Oceanographic data~---~www.aviso.altimetry.fr) gridded product is a dataset with horizontal resolution of 1/4$^\circ$ ($\sim$25~km) and temporal frequency of 1~day \citep{mertz2023aviso}. 
The dataset results from a Level~4 processing that combines data from multiple satellites~---~such as TOPEX/Poseidon, Jason-1, Jason-2 and Jason-3~---~to produce continuous fields of Absolute Dynamic Topography (ADT).
The altimeter exclusively measures surface height and geostrophic currents, thereby omitting subsurface shears and ageostrophic currents.
Surface current fields derived from ADT represent the geostrophic component of ocean currents, represented by (respectively in the eastward and northward directions)
\begin{equation}
u_g = -\frac{g}{f} \frac{\partial \text{ADT}}{\partial y}
\end{equation}
and
\begin{equation}
v_g = +\frac{g}{f} \frac{\partial \text{ADT}}{\partial x},
\end{equation}
where $f$ is the Coriolis parameter. 
Therefore, the SSalto/Duacs gridded product is an observational dataset, but it has some uncertainty associated with the gridding process~---~since along-track data are combined from different time instances to produce a continuous surface current field.

HYCOM (www.HYCOM.org) combines pressure coordinates, isopycnals and sigma coordinates  \citep{fox2002modular} which allows for a more accurate representation of ocean dynamics, particularly in areas with significant depth variations.
The model has fine spatial and temporal resolution and the ability to simulate complex ocean processes due to its hybrid approach, therefore including meso and submesocale effects not necessarily captured from altimeter data.
However, HYCOM would not have adequate resolution to resolve a considerable portion of the submesoscale with its effects incorporated into the current field through parameterizations \citep{GULA2022181}.
Here we employed the HYCOM NCODA GOFS 3.1, the most recent reanalysis of global HYCOM with 1/12$^\circ$ ($\sim$9~km) resolution and 41 vertical levels with temporal frequency of 3~hours.
The model resolves the primitive equations governing the conservation of momentum, mass and heat, while assimilating observational data on temperature, salinity and surface elevation.
HYCOM and most of the global forecasting models assimilate along-track altimetry, hence, part of the data observed in the gridded SSalto/Duacs is being incorporated by HYCOM during assimilation   \citep{chassignet2006ocean,chassignet2009us}.

The Ekman component represents the wind-driven part of surface currents, typically in the upper 50-100~meters.
Wind stress on the ocean surface creates a drag force that sets the water in motion, which diminishes with depth, following an exponential decay profile.
The GlobCurrent product (http://GlobCurrent.ifremer.fr/) merges altimetry data with wind fields to generate combined current fields, respectively from SSalto/Duacs maps and the wind stress from ERA~Interim \citep{rio}. 
The surface current fields include both the geostrophic component ($\mathbf{u}_g$) and the Ekman component ($\mathbf{u}_{ek}$) induced by the local wind shear stress
\begin{equation}
\mathbf{u}_{ek} = \beta(z) \mathbf{\tau} e^{i\theta(z)}, 
\end{equation}
where $z$ represents the vertical axis, $\beta(z)$ is vertical diffusivity, $\mathbf{\tau}$ is the wind stress and $\theta(z)$ is the direction of the Ekman current at different depths.
This provides a more comprehensive picture of the surface current field, accounting for both pressure-driven and wind-driven components.
With spatial and temporal resolutions of respectively 1/4$^\circ$ ($\sim$25 km) and 3~hours, the accuracy of GlobCurrent is considered high, particularly in regions well covered by satellite data \citep{cancet2019evaluation}. 
The most significant expected impact of adding the Ekman component is on the magnitude of the surface currents~---~the wind can enhance or weaken the geostrophic currents. 
The Ekman component can also influence the direction of surface currents, causing the overall current to veer slightly to the left (in the Southern Hemisphere) compared to the geostrophic current alone.
However, the position of mesoscale features (such as the dipole's central jet) is expected to be generally similar between ADT-derived geostrophic currents and its combination with the Ekman component.
Hence, since a significant portion of the current field comes from the geostrophy, it is expected that the GlobCurrent and SSalto/Duacs products will be very similar, with the addition of high frequency variability due to the Ekman component and its 3-hour resolution~---~in contrast to SSalto's daily resolution. 

Therefore, to summarize. 
SSalto/Duacs is observational data, HYCOM NCODA is a numerical model of primitive equations (with assimilated altimeter data) and GlobCurrent is a combination of observational data and numerical models.
HYCOM is more detailed and suitable for forecasts and dynamic studies, constrained however by the known limitations of any numerical model.
On the other hand, SSalto/Duacs and GlobCurrent data are limited to the geostrophic (and combined with Ekman) components of the surface currents for near real time and its historical database spanning over 25 years~---~they do not have a forecast product as those provided by global and regional numerical models.

\subsection{WW3 Wave Model}

\subsubsection{Idealized Simulations Setup}
The idealized simulations were conducted using WW3 version 7.14, with a spectral grid comprising 24~directions ($\Delta \theta =15^\circ$) and 32 frequencies, exponentially spaced from 0.0373~Hz to 0.7159~Hz with an increment factor of 1.1. 
The ST4 source term package \citep{st4} was implemented, configured with $\beta_{max}=1.55$, as described in \cite{ww3}, which extensively documents the model configuration, governing equations and computed bulk parameters~---~see also \cite{JK} for a discussion about different values of $\beta_{max}$ in the South
Atlantic Ocean.
The grid increment was 0.04$^\circ$ (4.4~km) in both $\Delta x$ and $\Delta y$, covering the area from 48$^\circ$W to 36$^\circ$W and 30$^\circ$S to 22$^\circ$S in the Southwestern Atlantic Ocean~---~see Figure~\ref{fig:fig9n}a~---~employing the ETOPO1 database \citep{etopo}.

Waves entered the domain from the southern boundary, about 100~km downwave of the dipole, and propagated northward ($\theta=180^\circ$). 
The simulations were conducted over a period of 30~days, allowing sufficient time for the system to reach a steady state within the computational domain.
Two initial states of WW3 simulations were considered, with $H_s$ set to 1~m and directional spreading ($\sigma$) to 15$^\circ$.
Peak wave periods ($T_p$) were assigned distinct values of 7~s and 15~s, resulting in corresponding steepness values ($\epsilon = \frac{H_s}{2} k_p$, where $k_p$ is the peak wavenumber) of 0.0822 and 0.0179, respectively.

Therefore, four different runs were analysed~---~two distinct values of $\epsilon$ and the surface current field.
Additionally, two control runs without current input were performed for both values of $\epsilon$.
Incorporating surface ocean currents into wave simulations reduces wave heights in most regions globally, primarily due to the reduction in relative wind caused by the alignment of winds and currents \citep{ECHEVARRIA2021101792}.
Here, to isolate the effects of currents on waves and simplify the analysis, wind input was excluded to eliminate continuous spectral modifications associated with wind forcing.
In our simulations, which were characterized by relatively low steepness values, dissipation and nonlinear interactions were considered negligible; however, they were activated in all cases to account for any potential influence on the results.

\subsubsection{Hindcast}
For the August and September 2010 hindcast, the WW3 version, grid and frequency discretization, configuration and bathymetry employed in the idealized simulations were applied.
However, two directional discretizations were employed, the original comprising 24~directions ($\Delta \theta =15^\circ$) and a higher resolution of 72~directions ($\Delta \theta=5^\circ$).
Eight runs were analyzed, using as input the three surface fields described in Section~\ref{3cur} and an additional control run with no current, for the two directional discretizations.
The wave model was also driven by hourly wind data from the ERA5 reanalysis dataset, produced by the European Centre for Medium-Range Weather Forecasts (ECMWF), with a spatial resolution of 0.25$^\circ$ (available at https://cds.climate.copernicus.eu/). 
Wave spectra forcing the model at the offshore boundaries were also from ERA5, with 24 directions, 30 frequency bands and a 0.5$^\circ$ grid with hourly temporal resolution.

\subsection{Altimeter data from the ESA Sea State Climate Change Initiative}

Until recently, the use of altimeter data often required preliminary low-pass filtering, which systematically resulted in the loss of small-scale features ($< 100$ km) \citep{Marechal-Ardhuin-2021}. 
In this context, the analysis of $H_s$ spatial variations was limited to wavelengths larger than 100~km, due to the noise associated with the tracking methods used to interpret 
altimeter waveforms \citep{Sandwell-etal-2014, fab}.
To overcome this limitation, the successful application of the Empirical Mode Decomposition (EMD) 
\citep{Huang-etal-1998} to the denoising of $H_s$ along-track series now makes it possible to investigate 
much smaller scales, possibly down to 15 km wavelength or less \citep{Quilfen-Chapron-2019}. 

In the present work, we use denoised $H_s$ \citep[see details in][]{Huang-etal-1998} from the European Space Agency (ESA) Sea State 
Climate Change Initiative (SeaState-CCI). In particular, the Level~3 (version~3) data is a global 
daily merged multi-sensor along-track satellite altimeter $H_s$ at about 6~km spatial resolution from 2002 to 2021 \citep{CCI-v3}.
From now on, we will refer to this dataset as CCI-Hs.

The CCI-Hs is based on a complete retracking of multiple satellite altimetry missions: 
Envisat, CryoSat-2, Jason-1, Jason-2, Jason-3, SARAL and Sentinel-3A. 
Taking into account that many altimeters are bi-frequency (Ku-C or Ku-S), 
for consistency reasons, only measurements in Ku band were used (except SARAL, where Ka band was used).
The selected tracks employed in the analysis that intersected the eddy dipole region~---~specifically, those tracks that crossed the dipole's central jet for the designated date~---~are listed in Table~\ref{tab:sat}.

Statistical metrics were computed to assess the agreement between altimeter measurements and \textit{in situ} data for each altimeter mission and year \citep{essd-12-1929-2020}.
The dataset is generated using a high-level denoising method, validation against in-situ measurements and numerical model outputs.
In the Sea State CCI version~3, the Jason-2 mission was selected as reference to inter-calibrate the remaining missions. 
A ﬁrst step was to calibrate this reference mission against \textit{in situ} data, considered as the ground truth. 
In a second step, the remaining missions (namely, Envisat, Jason-1, Saral, CryoSSat-2 and Jason-3) were compared against the calibrated Jason-2 data at
crossover locations in order to perform the inter-calibration. 
The analysis considered only altimeter and \textit{in situ} match-ups occurring more than 200~km from the coast, for the performance of calibrated and denoised altimeter significant wave height data. 
All missions exhibit a positive bias of less than 10~cm, while the root mean square error remains below 26~cm for all missions, corresponding to a normalized mean value below 11\%. 
Additionally, the scatter index is consistently lower than 9\%, while the coefficient of determination exceeds 0.96 for all missions, indicating strong agreement between the datasets.
A thorough discussion of the statistical metrics for each mission, including altimeter inter-calibration, long-term trends, cross-consistency analysis, quality assessment and the processing of altimeter data, is presented in \cite{essd-12-1929-2020}.

\begin{table}
\centering
\caption{List of selected tracks that crossed the eddy dipole region. Date is approximately at the center of the track segment.}
\begin{tabular}{|c|c|c|c|}
\hline 
\# & Date & Satellite & Cycle\tabularnewline
\hline 
\hline 
1 & 04-Aug-2010 00:47:00 & Jason-2 & 76\tabularnewline
\hline 
2 & 09-Aug-2010 13:11:00 & Jason-2 & 77\tabularnewline
\hline 
3 & 04-Sep-2010 12:36:00 & Envisat & 92\tabularnewline
\hline 
4 & 08-Sep-2010 07:07:00 & Jason-2 & 80\tabularnewline
\hline 
5 & 08-Sep-2010 01:01:00 & Envisat & 92\tabularnewline
\hline 
6 & 11-Sep-2010 01:07:00 & Envisat & 92\tabularnewline
\hline 
7 & 12-Sep-2010 16:41:00 & Jason-2 & 80\tabularnewline
\hline 
8 & 13-Sep-2010 06:10:00 & Jason-1 & 320\tabularnewline
\hline 
9 & 14-Sep-2010 01:13:00 & Envisat & 93\tabularnewline
\hline 
10 & 18-Sep-2010 05:05:00 & Jason-2 & 81\tabularnewline
\hline 
11 & 19-Sep-2010 12:39:00 & Cryosat-2 & 2\tabularnewline
\hline 
12 & 21-Sep-2010 12:37:00 & Cryosat-2 & 2\tabularnewline
\hline 
13 & 22-Sep-2010 14:40:00 & Jason-2 & 81\tabularnewline
\hline 
14 & 23-Sep-2010 12:35:00 & Cryosat-2 & 2\tabularnewline
\hline 
\end{tabular}
\label{tab:sat}
\end{table}

\section{Results}

\subsection{Idealized Simulations in the Vicinity of an Eddy Dipole} \label{ideal}
Figure~\ref{fig:fig9n}a shows the spatial variability of $H_s$, considering the 7~s waves.
The cyclonic eddy, with negative vorticity and clockwise rotation in the Southern Hemisphere, is more energetic than its anticyclonic pair with positive vorticity.
The refraction of the northward waves towards the central jet is hence asymmetric~---~the amount of bending from great circles is function of the ratio between the vertical component of the vorticity and group velocity (Equation~\ref{sec:mmodel:eq:ratioraycurv}).
The normalized relative vorticity is displayed in Figure~\ref{fig:fig9n}b, in the area shown as a black rectangle.

The relative difference in the $H_s$ field between runs with and without current is shown in Figure~\ref{fig:fig9n}, bottom row.
A maximum increase in $H_s$ for the 7~s waves of over 50$\%$ is observed in the central jet (Figure~\ref{fig:fig9n}c, around 43$^\circ$W and 25.75$^\circ$S), in the region between the maximum values of positive and negative vorticity, with maximum opposing surface currents of approximately 1~$ms^{-1}$.
Therefore, dipoles function as converging lenses for surface waves, directing their refraction toward the central jet.
For comparison, under these conditions, applying the linear theory to a 7~s monochromatic wave in the presence of homogeneous opposing currents predicts an increase of approximately 25$\%$ (Equation~\ref{sec:mmodel:eq:amplratio_phi}).
The combined effects of refraction and advection is responsible for a substantial increase in the wave energy and steepness over the central jet.
Similar spatial variability was observed for the 15~s waves (Figure~\ref{fig:fig9n}d), but with smaller intensification~~---~maximum of 33$\%$.
Comparatively, shorter period waves exhibit significant changes in both height and steepness, often resulting in energy dissipation through breaking.
Although not discussed in detail here, the directional spreading is also influenced by the dipole, as wave refraction alters their direction and consequently $\sigma$. 
With a narrow wave spectrum, refraction concentrate the wave energy into a tighter area, amplifying its impact.
On the other hand, if the wave spectrum is wide, the energy is distributed across a broader area since the various spectral components are more spread out.
The narrower the directional spreading, the more marked the alterations, exhibiting substantial broadening along the central jet.
Hence, the narrow $\sigma=15^\circ$ employed in the simulations is expected to cause a pronounced increase in energy.

Recent studies have demonstrated that refraction is the primary factor driving spatial variations of $H_s$ at mesoscale, between 20 and 100~km \citep{fab, romero2, Quilfen-etal-2018}.
More specifically, these gradients are primarily influenced by the rotational component of the surface current, whereas the divergent component has a negligible impact \citep{bia}.
That work identified a correlation between the spectral slopes of $H_s$ and kinetic energy of ocean currents, suggesting a link between wave energy distribution and current dynamics~---~the spectral slope of $H_s$ matches the spectral slope of the kinetic energy of the rotational component.
Spatial gradients of $H_s$ are more pronounced when the majority of the flow's kinetic energy is concentrated in the rotational component.

\begin{figure}[t!]
\centering
\includegraphics[width=1.0\textwidth]{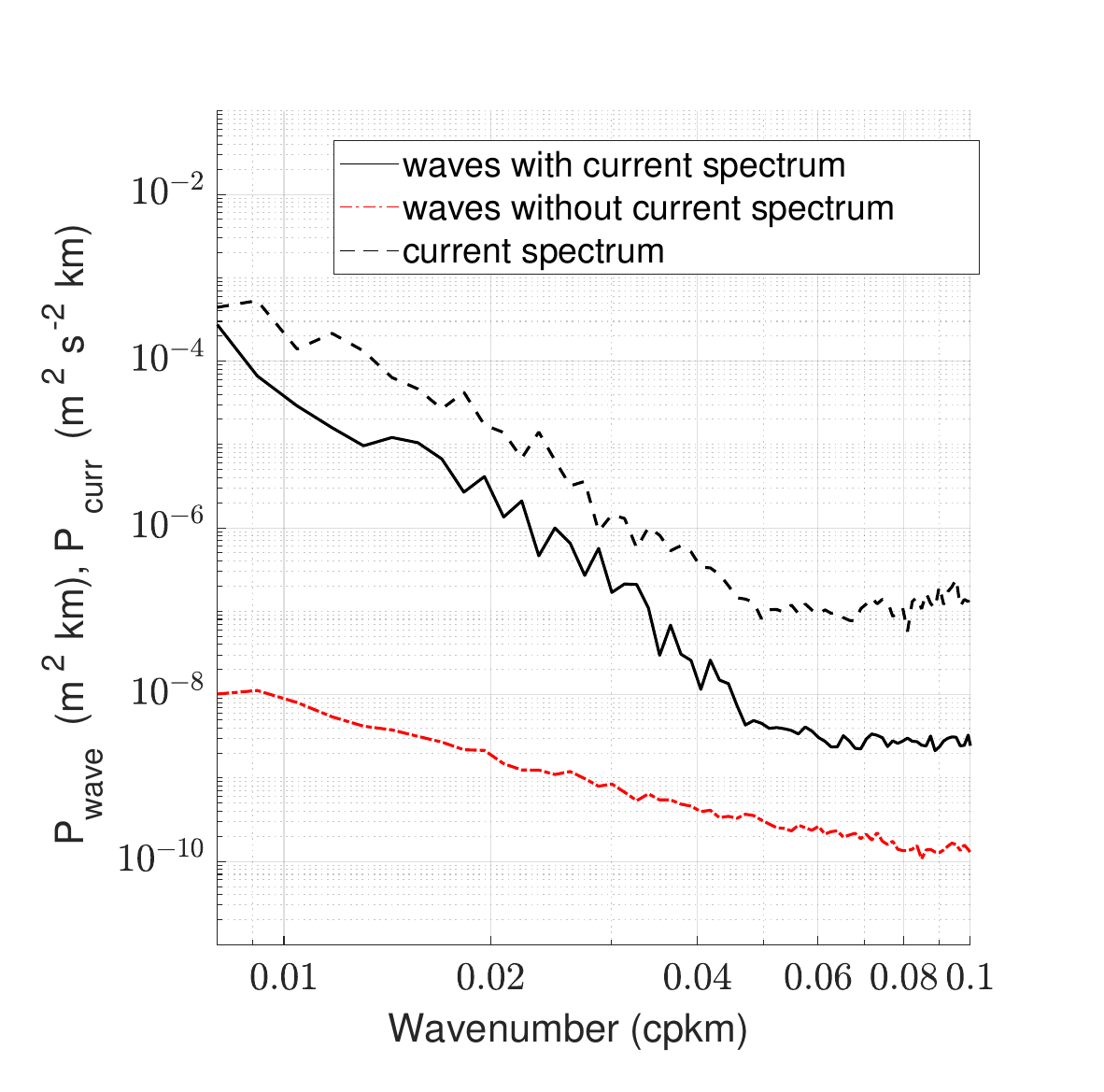}
\caption{\small{Averaged wavenumber spectra (in cycles per km) computed along the main axis~(E-W) in the black rectangle in Figure~\ref{fig:fig9n} for $H_s$ (solid line). The wavenumber $H_s$ spectrum without current input is displayed in red. Averaged wavenumber spectrum of surface current speed is the dashed line.}
\label{fig:fig9b}}
\end{figure}

The spatial variability in $H_s$ is therefore connected with the components of the flow, or more precisely with their amount of vorticity. 
The wavenumber spectra of spatial variations of currents and $H_s$ are shown in Figure~\ref{fig:fig9b}. 
Spectra were computed along the main axis (E-W) in the black rectangle depicted in Figure~\ref{fig:fig9n} during the whole period of simulation at the wave model resolution of 4.4~km (Nyquist wavelength of 0.11~cpkm), after detrending the data.
A Hann window was applied to the input matrix to minimize spectral leakage caused by discontinuities at the data boundaries. 
The window size was chosen to match the dimensions of the input matrix, using the same number of FFT elements as the original dataset, thereby preserving the spatial resolution of the analysis. 
To compensate for the energy attenuation inherent in windowing, the power spectral density was normalized by the total number of elements and the spatial resolution factors, ensuring an accurate representation of the total energy content in the wavenumber spectrum.

The $H_s$ spectra (solid line) between wavelengths 100 and 35~km show a trend with slope about $k^{-2.5}$ over most of the wavenumber range.
Beyond $\sim$0.05~cpkm in Figure~\ref{fig:fig9b} the slope of the spectra become flat, possibly caused by the effective spatial resolution of the models.
The current spectra is also shown in Figure~\ref{fig:fig9b}, as a dashed line, with slopes and trends similar to the $H_s$ spectra. 
The difference in the distribution of variance of $H_s$ spectra with and without current input across the spatial scales is expected to be expressive. 
The $H_s$ spectra without current input (the red dot-dashed line) is much less energetic and smoother in a region dominated by eddies.
In the idealized simulations, a unimodal spectrum propagating in deep water without surface currents is likely to exhibit a highly homogeneous distribution of $H_s$. 
At spatial scales of 100~km, which are typical for eddies in this region, the disparity in wave energy between runs with and without current input reaches its maximum. 
At smaller scales, the difference diminishes progressively, as smaller eddies exhibit lower energy levels~---~Figure~\ref{fig:fig9n}b.

\subsection{Ocean Eddy Dipole in the Southwestern Atlantic Ocean}
We now focus our attention on a two-month hindcast with a well developed dipole in the Southwestern Atlantic Ocean.
Among the various tracks traversing the region, we selected only those in which the central jet maintained an approximately north–south orientation and with mainly southward currents, typically in the opposite direction of swell \citep{nvc}.
Under these circumstances, the current effects on waves are expected to cause the maximum amplification of $H_s$, as discussed in Section~\ref{back}.
The CCI-Hs data were employed to assess the response of the wave model to the different current inputs, listed in Table~\ref{tab:sat}.

The South Brazilian Bight (SBB) is a geographical feature located off the coast of southern Brazil and known for the recurrent presence of ocean eddy dipoles (see a schematic representation in Figure~\ref{fig:fig1aa}). 
Within the meanders, the current can spawn swirling eddies, 
whose size ranges from a few kilometers to tens of kilometers and can persist for weeks or even months \citep{arruda2019dipole}.
The ocean circulation at the Sao Paulo plateau was addressed by \cite{Belo2011} based on a series of measured current  and hydrographic transects, indicating a pattern of successive eddy structures. 
According to \cite{Andrioni2012}, dipoles are recurrent features as its surface signature can be frequently observed in satellite derived sea surface height in the region. 
Moreover, the Sao Paulo plateau acts as a natural eddy corridor where dipoles gradually move southwestward \citep{thiago}.  

\begin{figure}[t!]
\centering
\subfigure[]{%
\label{subfig:1aaa}\resizebox*{6.5cm}{!}{\includegraphics{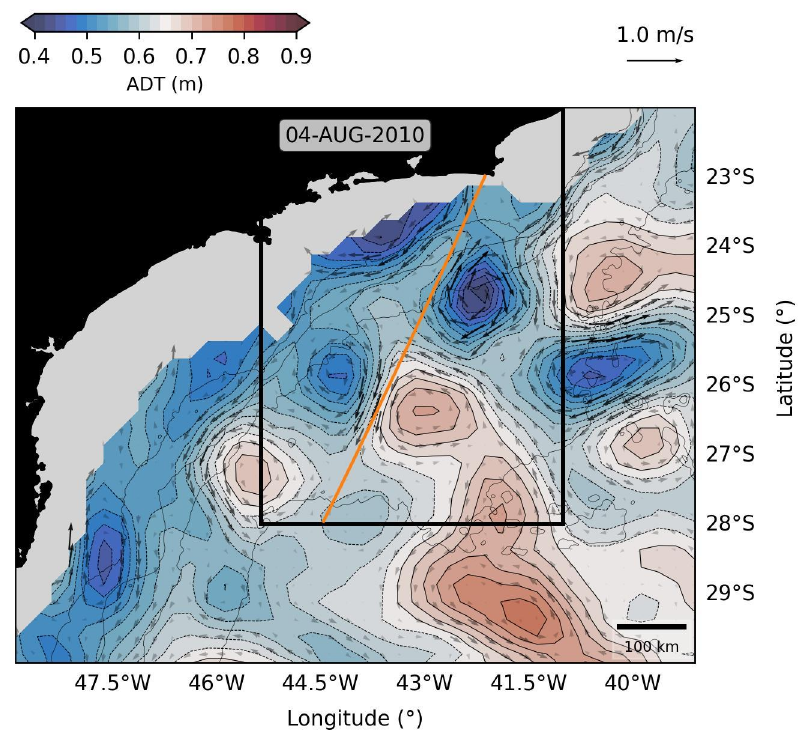}}}
\subfigure[]{%
\label{subfig:1bbb}\resizebox*{6.5cm}{!}{\includegraphics{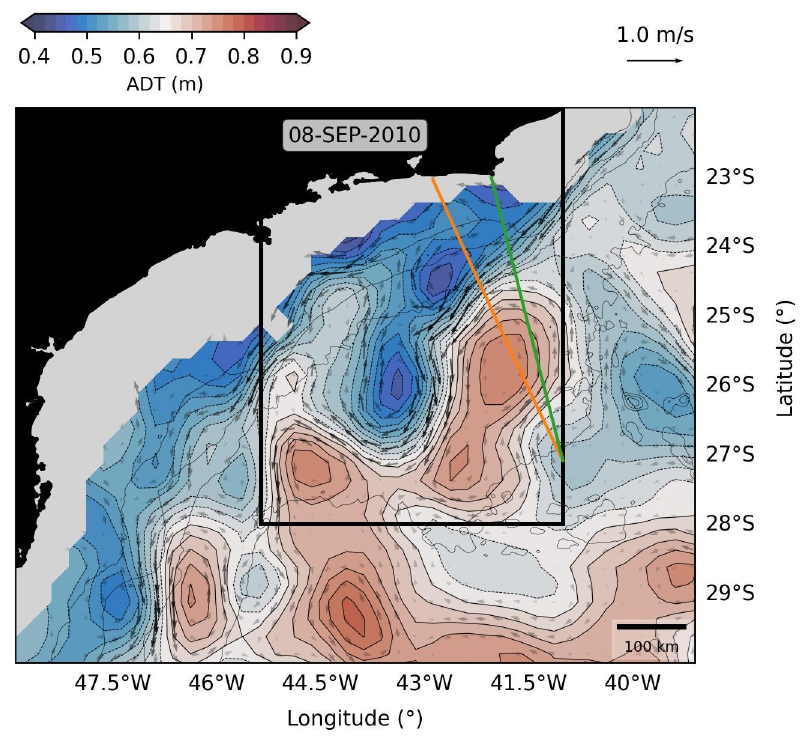}}}
\subfigure[]{%
\label{subfig:1ccc}\resizebox*{6.5cm}{!} {\includegraphics[width=\textwidth]{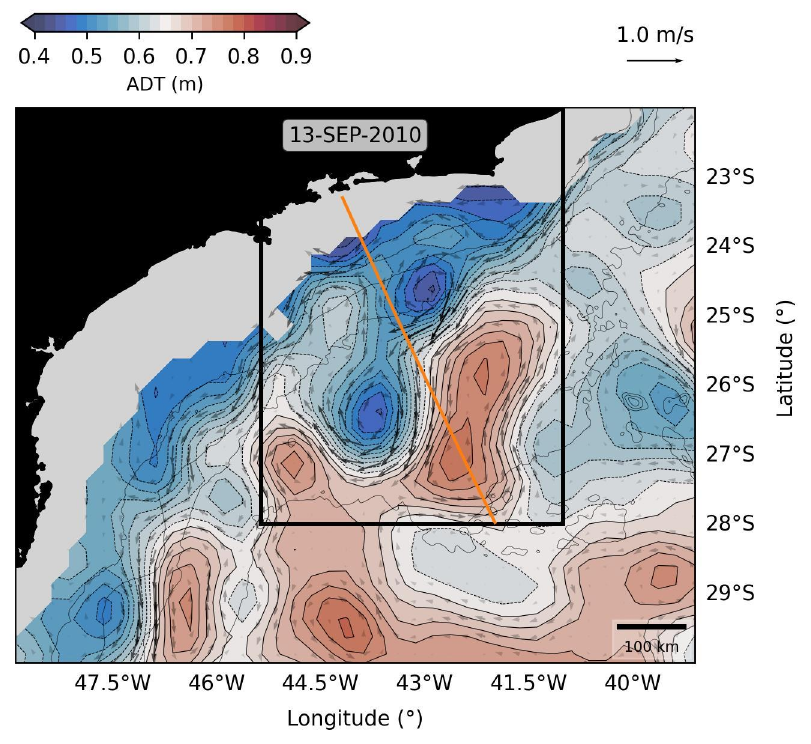}}}
\subfigure[]{%
\label{subfig:1ddd}\resizebox*{6.5cm}{!} {\includegraphics[width=\textwidth]{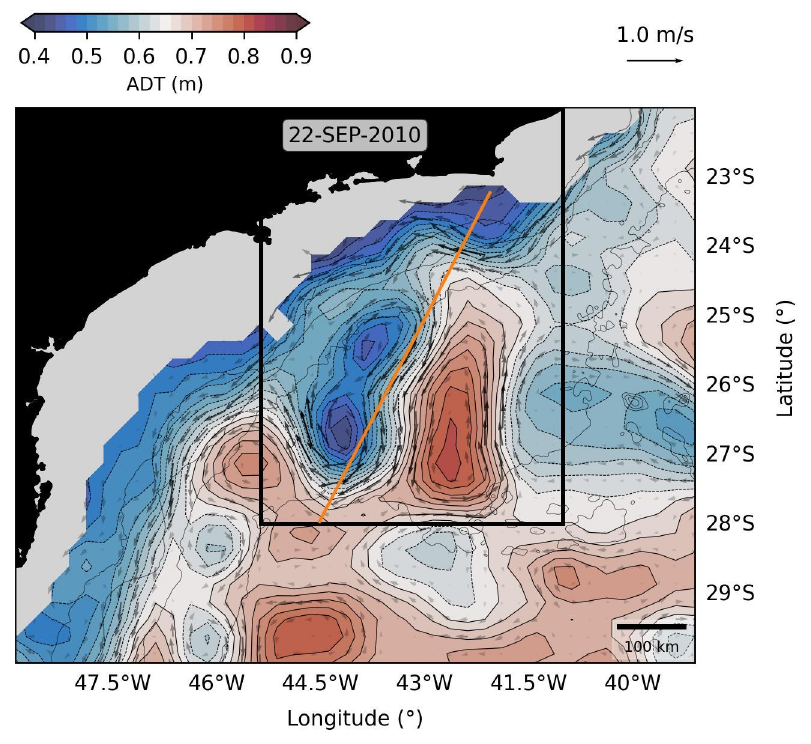}}}
\caption{\small{Fields of Absolute Dynamic Topography (ADT) from SSalto/Duacs and the derived geostrophic currents superimposed with the tracks shown as straight colored lines. [abcd] are four out of the 14 cases during the period of August and September 2010 that the satellite tracks crossed the dipole's central jet, respectively tracks: \#1; \#4 and \#5; \#8; \#13 (Table~\ref{tab:sat}). The scale of the vectors is displayed on the top right corner}.}
\label{fig:fig1bb}
\end{figure}

The time evolution of some of the delayed-time merged ADT fields and derived absolute surface geostrophic velocities, from the SSalto/Duacs multimission altimeter data, are shown in Figure~\ref{fig:fig1bb}.
The surface geostrophic currents from the ADT fields are one of three distinct input fields for the wave model. 
The low and high features that make up the asymmetric eddy dipole are highlighted within the black rectangle, with the satellite tracks superimposed.
The side-by-side high and low ADT values characterize an eddy dipole, with adjacent anticyclonic and cyclonic eddies, respectively.

\begin{figure}[t!]
\centering
\subfigure[]{%
\label{subfig:1a}\resizebox*{6.5cm}{!}{\includegraphics{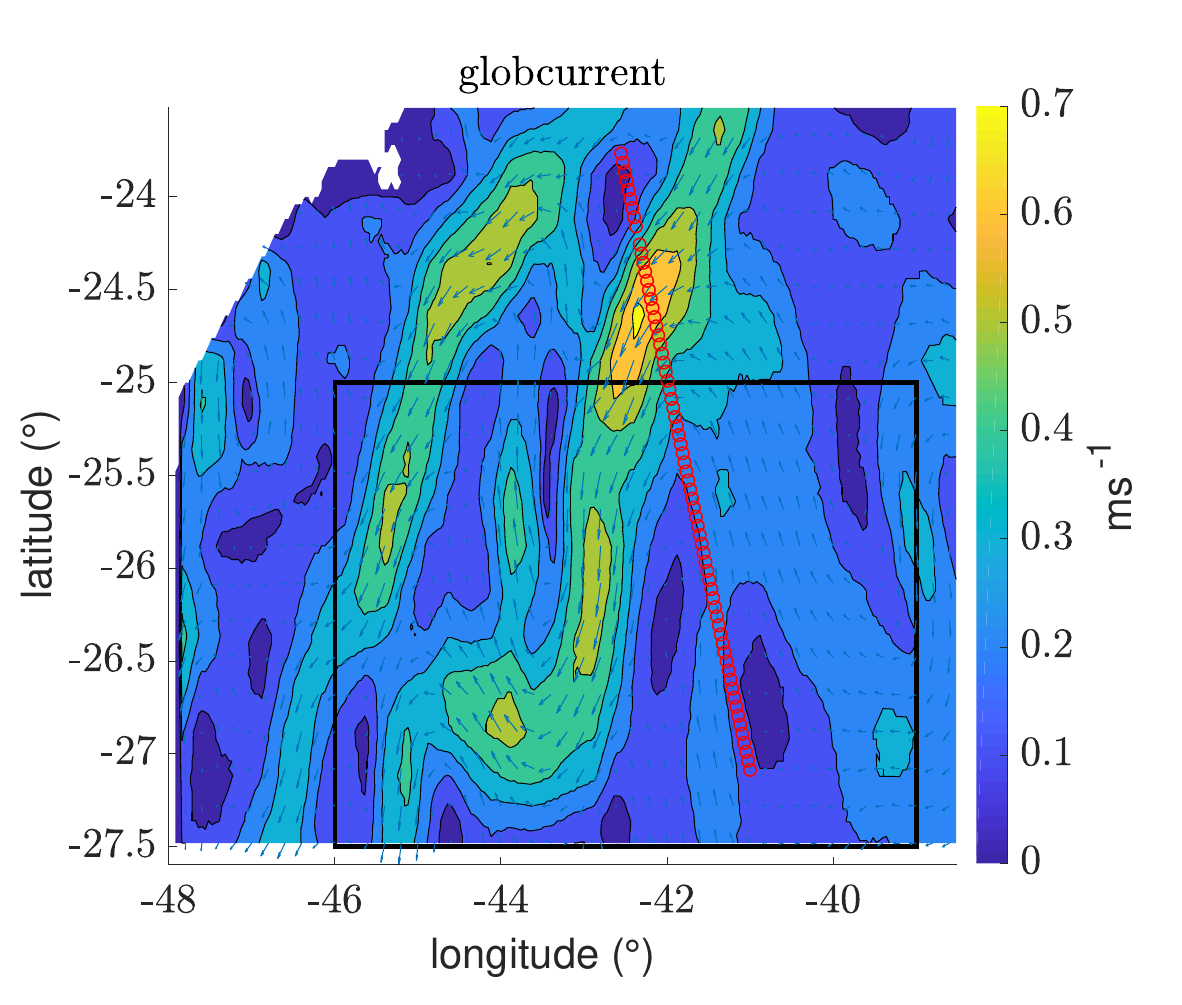}}}
\subfigure[]{%
\label{subfig:1b}\resizebox*{6.5cm}{!}{\includegraphics{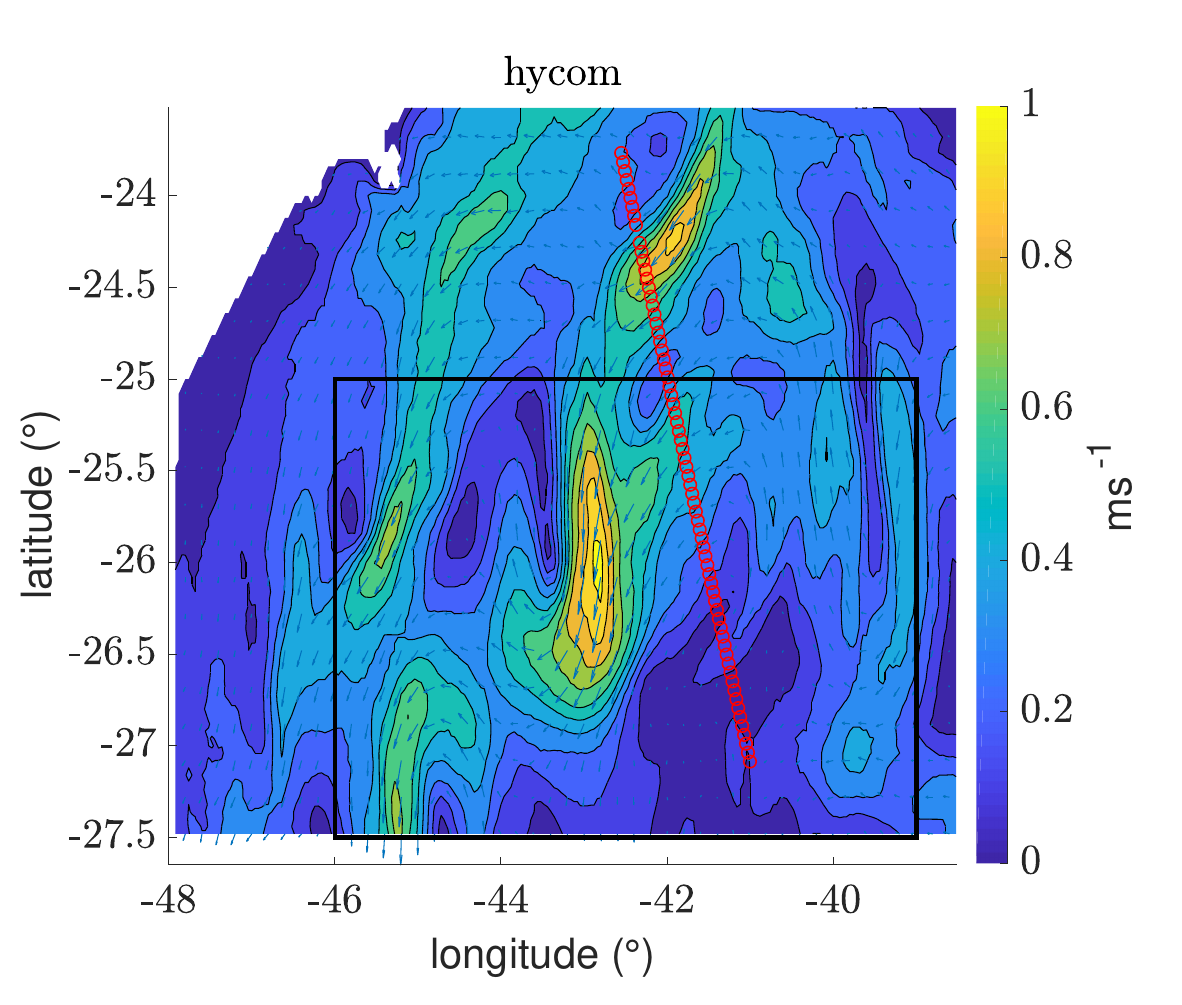}}}
\subfigure[]{%
\label{subfig:1c}\resizebox*{6.5cm}{!} {\includegraphics[width=\textwidth]{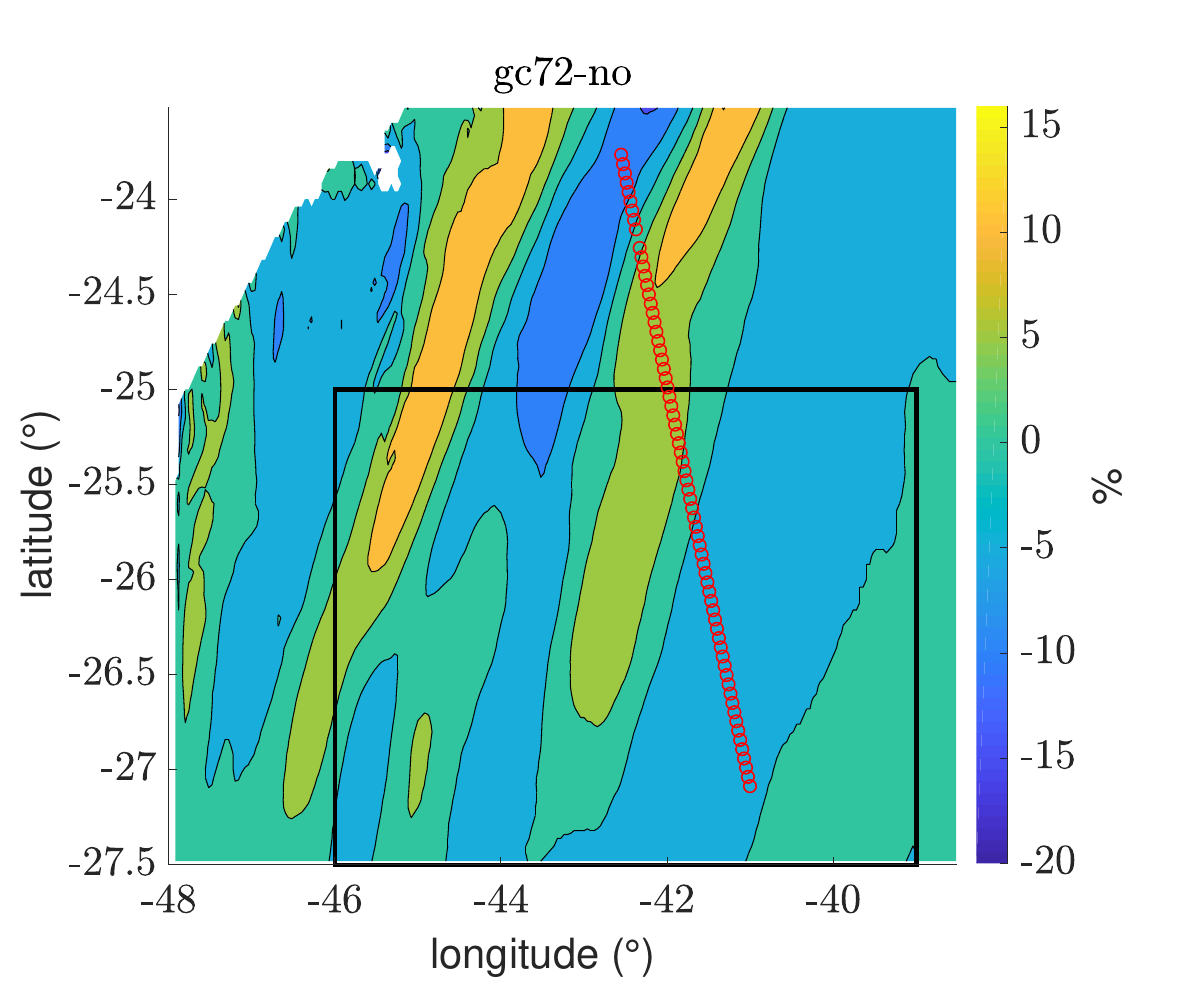}}}
\subfigure[]{%
\label{subfig:1d}\resizebox*{6.5cm}{!} {\includegraphics[width=\textwidth]{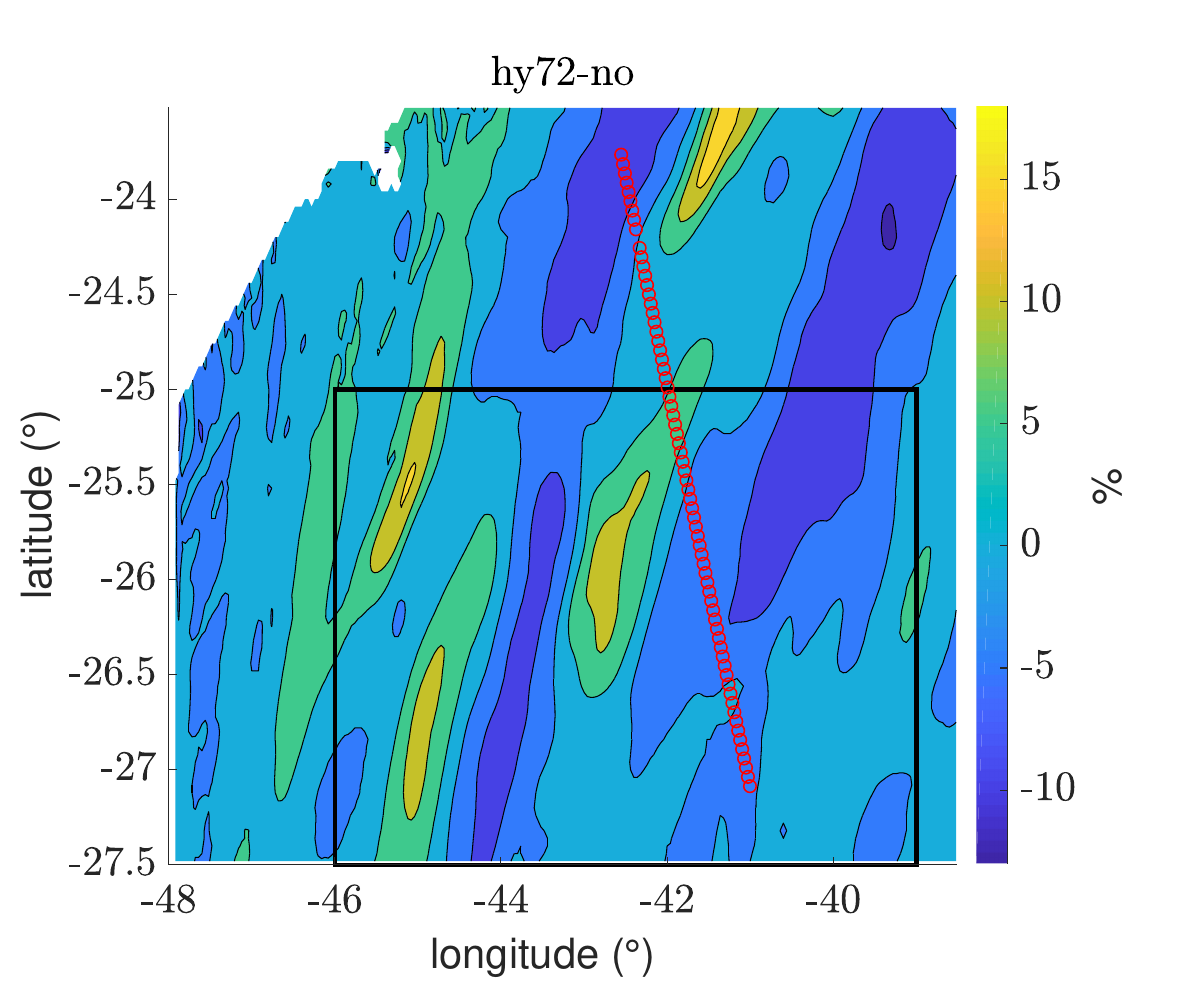}}}
\caption{\small{Example of one event that HYCOM misplaced the position of the strongest currents associated with the dipole~---~see also Figure~\ref{fig:fig4}. Top row: surface current field from GlobCurrent (left) and HYCOM (right) for 08-Sep-2010 at 07:00:00. The arrows indicate the current direction and magnitude sub-sampled every four bins. Red circles are the ground track of Jason-2 cycle~80. Bottom row: $H_s$ relative difference between runs with and without current for GlobCurrent (left) and HYCOM (right). The results employing the SSalto current field are similar to GlobCurrent.}
\label{fig:fig1}}
\end{figure}

Figure~\ref{fig:fig1}, top row, shows the other two input current fields employed in the wave model runs, respectively GlobCurrent and HYCOM, for satellite track $\#$4 in Table~\ref{tab:sat}~---~see also the ADT field in Figure~\ref{fig:fig1bb}b.
The current fields shown in Figure~\ref{fig:fig1} differ in magnitude and position of the maximum values, reflected in the relative difference in $H_s$ between runs with and without current input displayed in the bottom row.
The HYCOM currents are stronger than GlobCurrent (as generally expected) since the ageostrophic component is included, hence the relative differences in $H_s$ are slightly more pronounced.
HYCOM in this event misplaced the position of the central jet, even though the SSH field derived from altimeter data was incorporated during assimilation~---~the same altimeter data employed in the processing of SSalto and GlobCurrent.

\begin{figure}[t!]
\centering
\subfigure[]{%
\label{subfig:4a}\resizebox*{6.5cm}{!}{\includegraphics{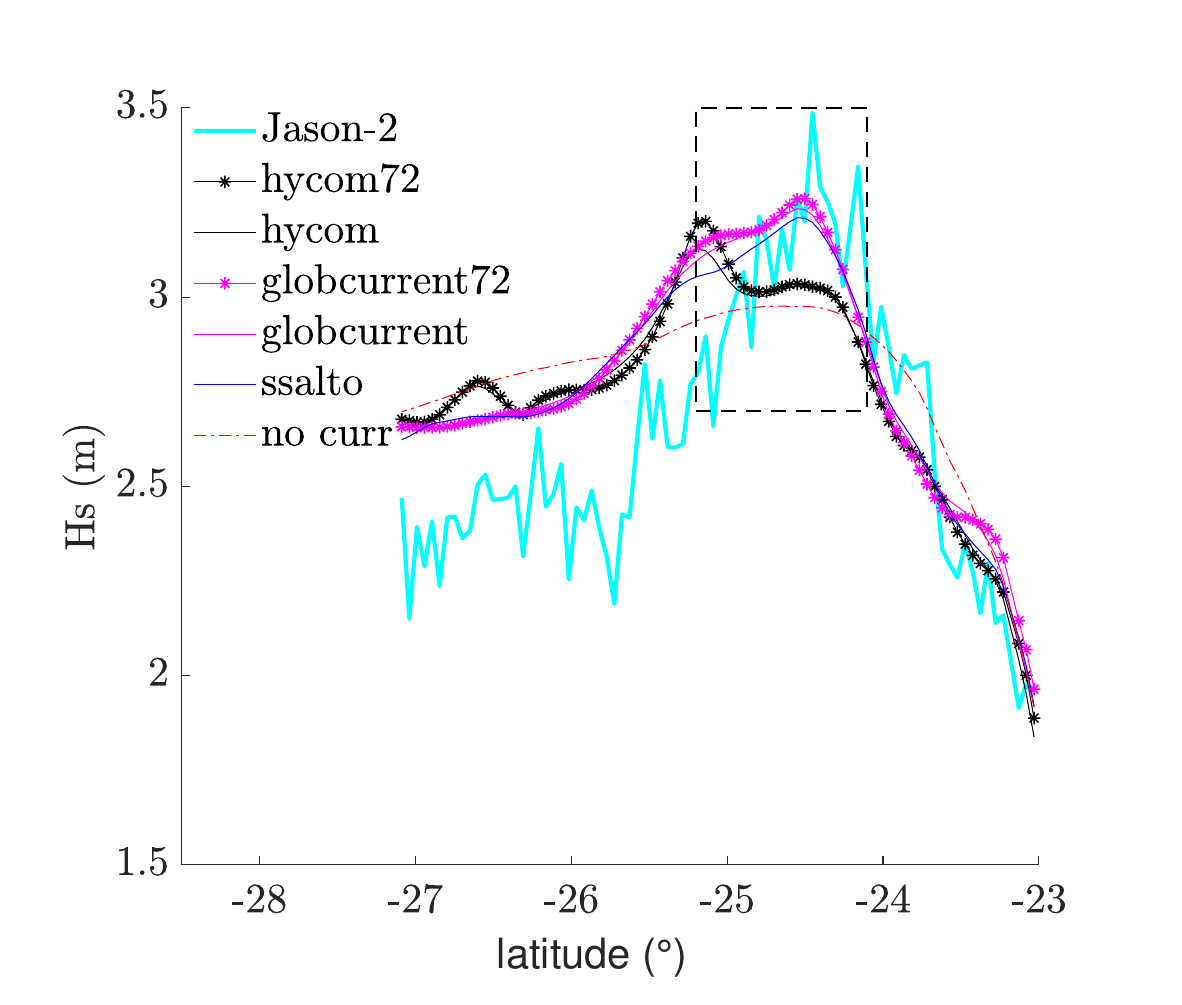}}}
\subfigure[]{%
\label{subfig:4b}\resizebox*{6.5cm}{!} {\includegraphics[width=\textwidth]{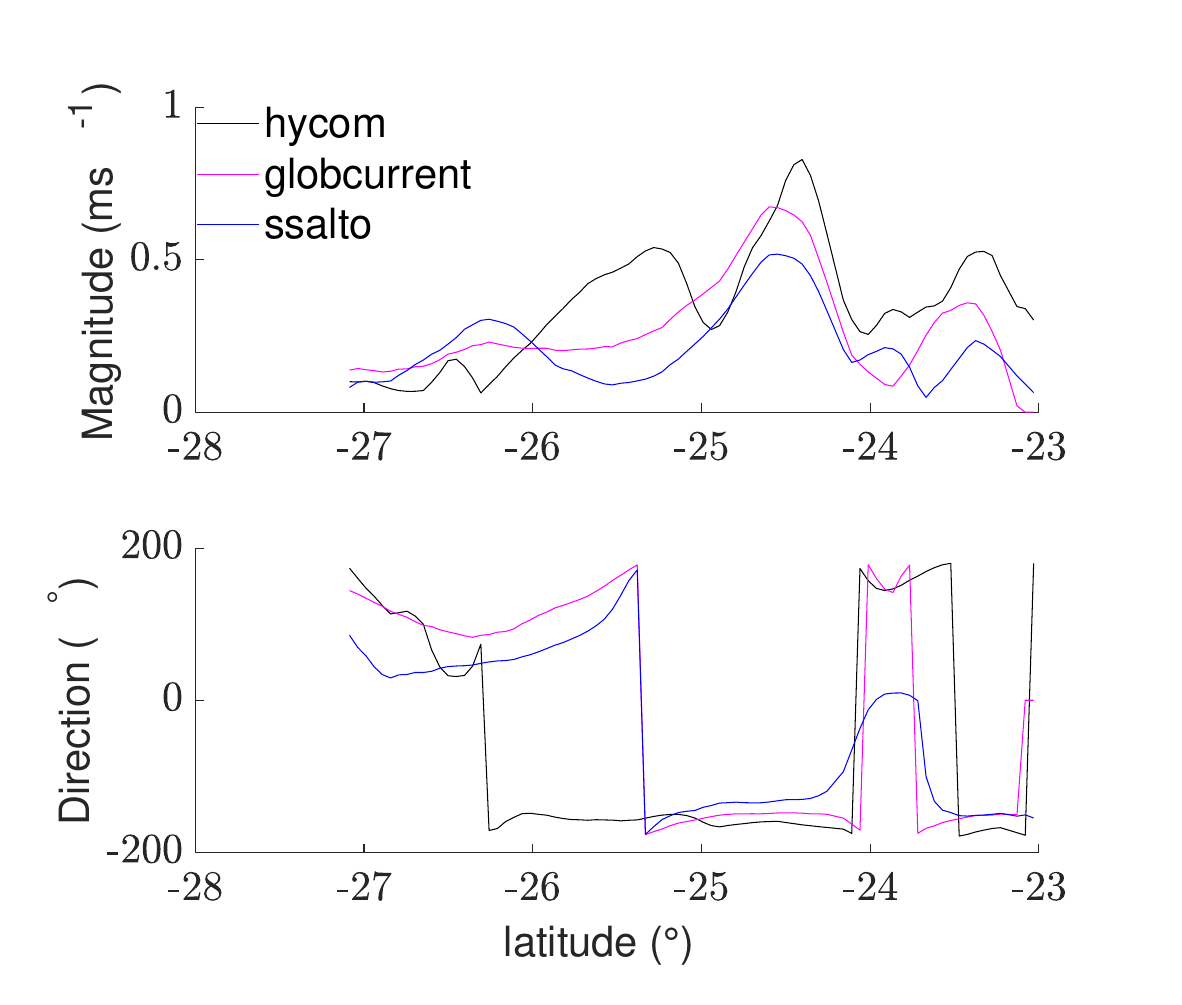}}}
\caption{\small{On the left, $H_s$ along the Jason-2 track (for 08-Sep-2010 at around 07:07:00) including different wave model runs. The dashed rectangle is the region that the track crossed the dipole's central jet. The ADT and derived geostrophic currents  superimposed with the satellite track crossing the eddy dipole are shown in Figure~\ref{fig:fig1bb}b. On the right, magnitude (top) and direction (bottom) of the surface currents used as input for the wave model, interpolated along the track. Statistical parameters were also computed for each of the 14 track segments that crossed the dipole~---~see Table~\ref{tab:table1}.}
\label{fig:fig4}}
\end{figure}

$H_s$ along the Jason-2 track are presented in Figure~\ref{fig:fig4} (left) with the several runs of the wave model.
Jason-2 measured a $H_s$ maximum value of 3.5~m at 24.45$^\circ$S, well spotted by SSalto and GlobCurrent runs (around 3.2~m) but missed in the simulation using the HYCOM input.
SSalto and GlobCurrent currents, interpolated along the satellite track (shown in Figure~\ref{fig:fig4} top right and bottom right, respectively magnitude and direction), indicate both the same position of maximum values at 24.40$^\circ$S.
The addition of the Ekman component in GlobCurrent increased its magnitude but kept roughly the same position depicted in SSalto.
HYCOM, however, misplaced the position of the maximum value by $\sim$20-30~km.
Although the magnitude is higher in HYCOM along the ground track, the simulated direction also differ from the ADT-derived currents.
SSalto and Globcurrent are roughly south-southwestward, in the opposite direction of the north-northeastward swell, whereas HYCOM presents a more (orthogonal) westward component, hence with less influence on the wave propagation (Equation~\ref{sec:mmodel:eq:doppler}).

\begin{figure}[t!]
\centering
\subfigure[]{%
\label{subfig:4a}\resizebox*{6.5cm}{!}{\includegraphics{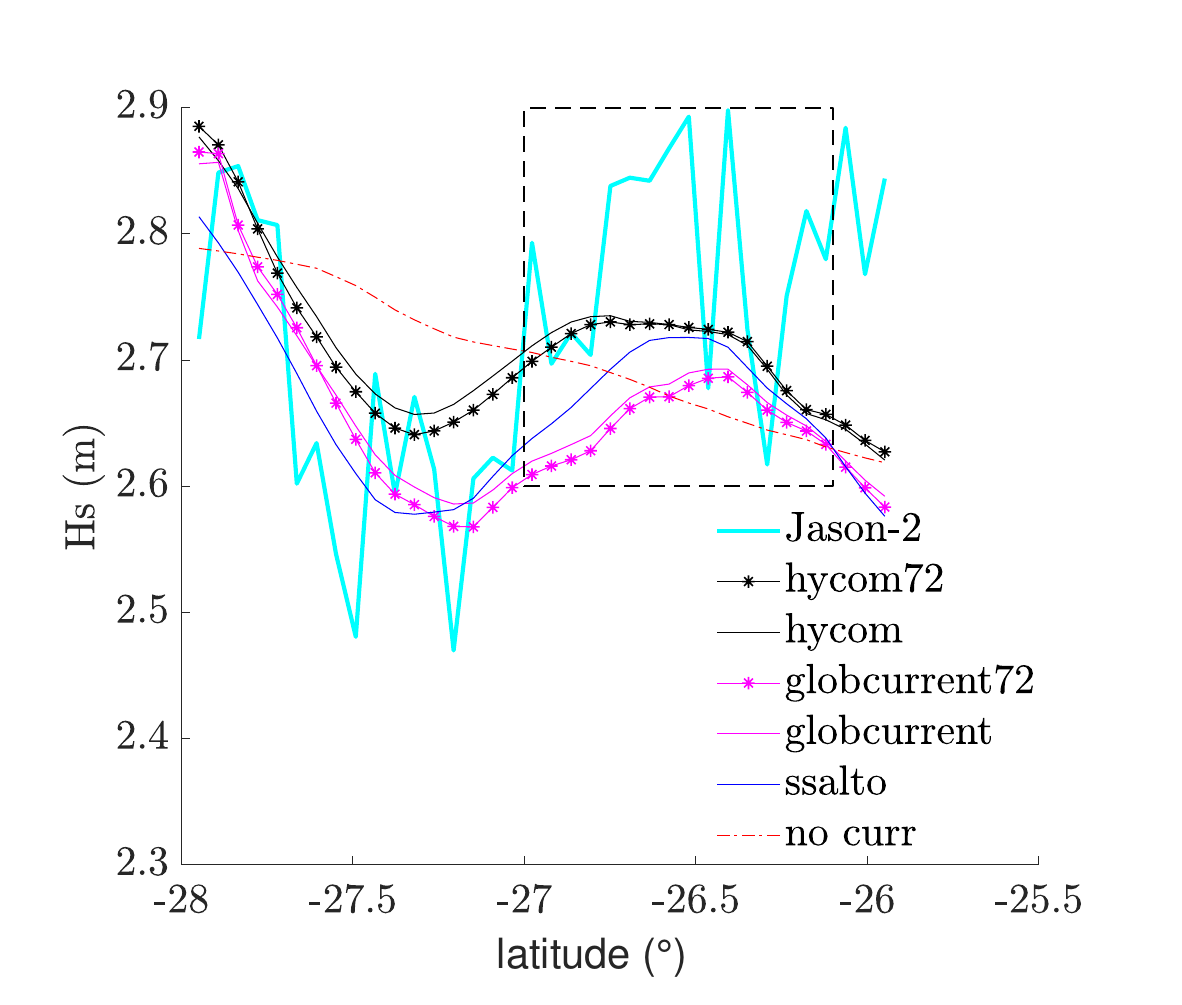}}}
\subfigure[]{%
\label{subfig:4b}\resizebox*{6.5cm}{!} {\includegraphics[width=\textwidth]{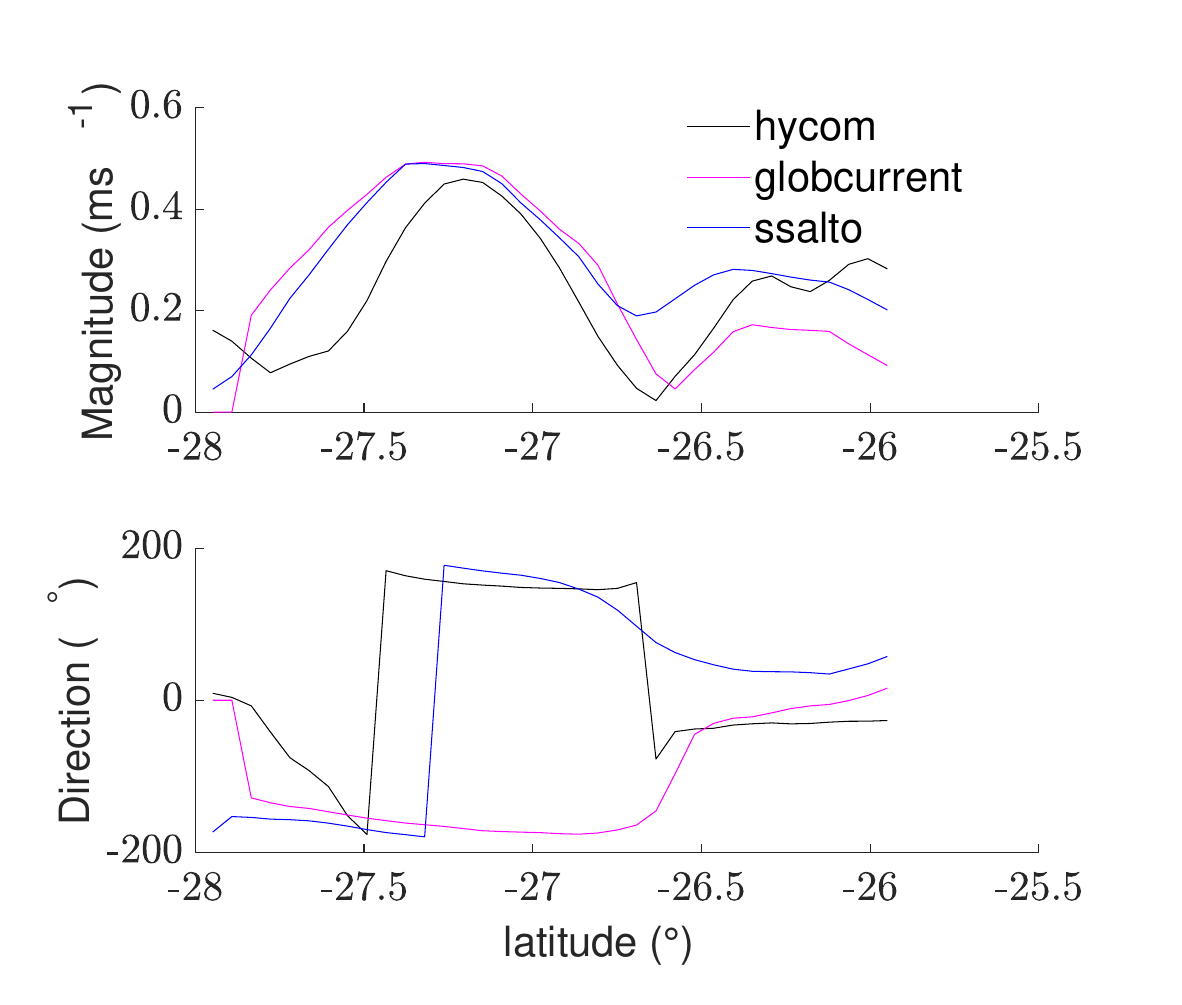}}}
\caption{\small{Same as Figure~\ref{fig:fig4}, for 22-Sep-2010 at around 14:40:00.}
\label{fig:fig4n}}
\end{figure}

Another example of the misplacement of the central jet is illustrated in Figure~\ref{fig:fig4n}, which can also be observed in the ADT field for satellite track~\#13 in Figure~\ref{fig:fig1bb}d (see Table~\ref{tab:sat} for reference).
The current fields exhibit discrepancies in both magnitude and spatial positioning of their maximum values, which are reflected in the relative differences in $H_s$ between simulations with and without current input. 
The left panel presents $H_s$ measurements along the Jason-2 track, comparing them with multiple wave model runs. 
Jason-2 recorded a maximum $H_s$ value of 2.9~m at 26.41°S, which was accurately captured by the SSalto and GlobCurrent runs (with peaks around 2.7~m) but was missed by the simulation using HYCOM input. 
The SSalto and GlobCurrent current fields, when interpolated along the satellite track (shown in Figure~\ref{fig:fig4n}, top right and bottom right, representing magnitude and direction, respectively), consistently identified the maximum value at 26.41°S. 
In contrast, HYCOM misplaced the position of the maximum value by approximately 30~km.

\begin{table}[t]
\centering
\caption{Mean statistical parameters of $H_s$ for the 14 selected ground tracks listed in Table~\ref{tab:sat}. In black, the segments that crossed the dipole region (including the central jet) with a total of 971 points. In red, the segments that crossed the central jet only, with a total of 476 points. Respectively, CORR is the Pearson Correlation Coefficient, bias between model runs and altimeter, Root Mean Squared Error (RMSE), Scatter Index (SI) and Variance (VAR). HY72 is the run with the HYCOM surface current used as input for WW3 with a resolution of 72~directions, while HY if for 24~directions. GC72 is the run with the GlobCurrent surface current used as input for WW3 with a resolution of 72~directions, while GC is for 24~directions. SS is the run with the  SSalto DUACS surface current used as input for WW3 with a resolution of 24~directions. NOC is the WW3 run without surface current as input}.
\begin{tabular}{|c|c|c|c|c|c|}
\hline 
 & CORR & BIAS~(m) & RMSE~(m) & SI~(\%) & VAR~($m^2$)\tabularnewline
\hline 
\hline 
HY72 &  0.64 \textcolor{red}{0.55} & 0.059 \textcolor{red}{0.049} & 0.26 \textcolor{red}{0.26} & 10.5 \textcolor{red}{10.1} & 0.076 \textcolor{red}{0.032} \tabularnewline
\hline 
HY & 0.64 \textcolor{red}{0.57} & 0.062 \textcolor{red}{0.043} & 0.26 \textcolor{red}{0.26} & 10.4 \textcolor{red}{9.9} & 0.072 \textcolor{red}{0.025}\tabularnewline
\hline 
GC72 & 0.66 \textcolor{red}{0.61} & 0.039 \textcolor{red}{0.026} & 0.24 \textcolor{red}{0.24} & 9.8 \textcolor{red}{9.2} & 0.062 \textcolor{red}{0.026}\tabularnewline
\hline 
GC & 0.67 \textcolor{red}{0.61} & 0.041 \textcolor{red}{0.023} & 0.24 \textcolor{red}{0.23} & 9.8 \textcolor{red}{9.0} & 0.060 \textcolor{red}{0.021}\tabularnewline
\hline 
SS & 0.68 \textcolor{red}{0.65} & 0.040 \textcolor{red}{0.022} & 0.23 \textcolor{red}{0.22} & 9.4 \textcolor{red}{8.6} & 0.057 \textcolor{red}{0.022}\tabularnewline
\hline 
NOC & 0.53 \textcolor{red}{0.30} & 0.040 \textcolor{red}{-0.008} & 0.25 \textcolor{red}{0.23} & 9.9 \textcolor{red}{8.9}& 0.045 \textcolor{red}{0.006}\tabularnewline
\hline 
\end{tabular}
\label{tab:table1}
\end{table}

The primary reason for the greater mismatch of the eddy dipole simulated by global HYCOM, compared to derived gridded altimetry, is that the reanalysis assimilates along-track sea level anomalies. 
Although the reanalysis offers higher spatial resolution, the available tracks for assimilation cover smaller parts of the domain and may not have sampled the eddy dipole, thereby reducing the impact of data assimilation on correcting the position of this particular mesoscale feature. 
Additionally, all oceanic reanalyses contain other sources of error that contribute to deviations of global numerical simulations from the observed ocean state at regional scales. 
These include the lack of subsurface observations for assimilation, inaccuracies in topography representation, uncertainties in initial and boundary conditions, surface forcing, and limitations in the parameterizations used for sub-grid processes \citep{Lima2019, deSouza2020, thiago, Trott2023}.

It is clear in Figure~\ref{fig:fig4} and Figure~\ref{fig:fig4n} (left) that tripping the wave model resolution from 24 to 72~directions has a marginal impact on $H_s$, despite the substantial increase in computing time.
That is also evident considering the statistical parameters of the selected 14~tracks.
The numbers in black in Table~\ref{tab:table1} correspond to the ground track segments that crossed the dipole region, with Figure~\ref{fig:fig1} and Figure~\ref{fig:fig4} as examples of the length of the segments of approximately 400~km each.
These segments cross not only the dipole but the adjacent area, where the current field is less intense and hence less relevant for the wave field.
The numbers in red correspond to the track segments of the much smaller central jet region, depicted as the dashed rectangle in Figure~\ref{fig:fig4}.
As expected in such a dynamically complex region, correlation with the altimeter $H_s$ considering the run with no current (dash-dotted red line in Figure~\ref{fig:fig4}) is low but with comparatively small RMSE, since the altimeter data go up and down relative to the much flatter no current $H_s$ line.
Results improve when considering the runs with currents as input, specially for SSalto and GlobCurrent (the latter with slightly poorer results).
HYCOM surface currents, despite being physically more sound with the inclusion of the ageostrophic component as input for the wave model, has worse results caused by the greater difficulty of placing the currents on their correct positions.
Both ADT-derived surface current inputs cause similar impacts on the wave field, with no substantial improvement with the inclusion of the Ekman component.
Considering the dipole and its adjacent area (numbers in black), SSalto presents slightly better results.
That is more evident considering the much more difficult task of placing correctly in space the comparatively small region of the central jet (numbers in red).
The correlation coefficient is a good indicator whether the $H_s$ curves move together consistently, with the SSalto current input presenting a correlation coefficient of 65$\%$ with the altimeter data, ten points greater than HYCOM.
Additionally, the biases observed in HYCOM simulations are twice as large as those in the SSalto simulations~---~however, the differences in RMSE are not statistically significant.
HYCOM is particularly effective in capturing a broad range of dynamical processes, including ageostrophic and smaller-scale components essential for accurate ocean energy representation. 
However, discrepancies with gridded SSalto/Duacs data may arise from potential assimilation inaccuracies and inherent model limitations, such as parameterization errors and uncertainties in atmospheric forcing.

\begin{figure}[t!]
\centering
\includegraphics[width=1.0\textwidth]{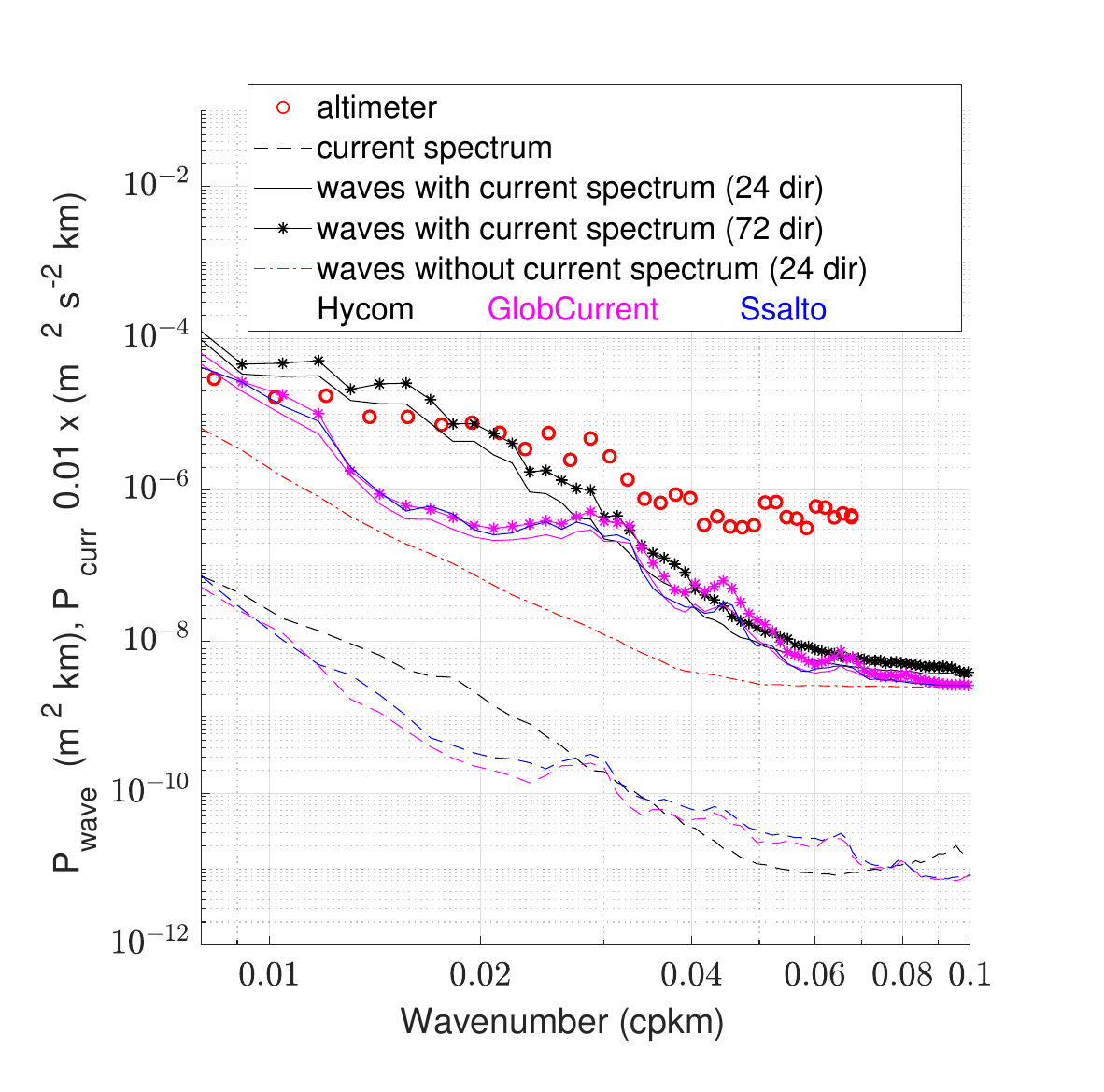}
\caption{\small{Wavenumber spectra (in cycles per km) computed along the main axis (E-W) in the black rectangle in Figure~\ref{fig:fig1} for $H_s$ (solid lines) averaged for the month of September. The wavenumber $H_s$ spectrum without current input is displayed as the dash-dotted red line. For comparison, the along-track averaged spectrum corresponding to all tracks listed in Table~\ref{tab:sat} are presented as red circles. Averaged wavenumber spectra of surface current speed are the dashed lines~---~scaled by a factor $10^{-2}$ for easy of visualization.}}
\label{fig:fig5}    
\end{figure}



Wavenumber spectra were computed for the month of September, shown in Figure~\ref{fig:fig5}, following the same procedure described in Section~\ref{ideal}~---~solid lines for $H_s$ and dashed lines for current speeds.
In the range from 100 to 30~km, the current spectrum from HYCOM is more energetic (dashed black) than SSalto (dashed blue) and GlobCurrent (dashed magenta), which show similar patterns.
The difference in energy at the lower wavenumber band between current spectra from HYCOM and from the ADT-derived field could be partially attributed to the ageostrophic component, not present in SSalto and GlobCurrent (although the latter includes the Ekman component).
The surface current grid derived from the composition of several altimeter tracks with a course time resolution of 24~hours is expected to be smoother, which could also add to the discrepancy in spectral energy.
The surface currents were interpolated to match the wave model’s resolution of 4.4~km. 
However, SSalto and GlobCurrent datasets have a coarser spatial resolution of 25~km, compared to HYCOM’s 9~km. 
Consequently, interpretations of the spectral behaviour beyond $\sim$0.02~cpkm~---~specially for the ADT-derived currents~---~should be approached with caution.

The slopes of the Hs spectra (solid lines) are similar to the current spectra, with the HYCOM currents as input (solid black) also exhibiting higher spatial variability compared to SSalto (solid blue) and GlobCurrent (solid magenta). 
Notably, there are striking differences in the spectral content between the runs with and without current input (dash-dotted red line). 
Tripping the directional spectrum results in a minor increase in the spectral content. 
Similar to the current spectra, all the Hs spectra converge at approximately 0.03~cpkm. 
The averaged along-track spectrum for all tracks listed in Table~\ref{tab:sat} is presented as red circles, illustrating the spatial variability of $H_s$ along the altimeter tracks. 
The wavenumber spectrum of $H_s$ shows strong agreement with our model results, particularly with simulations that incorporate ocean current effects from HYCOM. 
At wavelengths shorter than 50~km, the altimeter data are influenced by high-frequency noise, resulting in a nearly flat spectrum for wavenumbers greater than 0.02~cpkm. 
Although comparing the $H_s$ spectra generated from different current fields used as input to the wave model is not straightforward, it offers an interesting opportunity to infer distinct responses of the wave field to the various components of the surface currents. 
The surface current from HYCOM is expected to be more energetic, not only due to the presence of the ageostrophic component but also because of the comparatively less smoothed currents. 
Consequently, the $H_s$ spectrum using the HYCOM input is also expected to be more energetic at lower wavenumbers.

\section{Summary and Conclusions}

We investigate the influence of surface currents on the spatial variability of the wave field in the vicinity of an ocean eddy dipole. 
Eddy dipoles are ubiquitous mesoscale features in the oceans, characterized by adjacent cyclonic and anticyclonic, counter-rotating eddies separated by a central jet. 
These structures can form in regions dominated by intense currents as well as in areas characterized by relatively weak flow regimes.
An idealized setup using the WW3 wave model was employed to analyze the impact on significant wave height ($H_s$). 
Additionally, a two-month hindcast of an intense dipole event in the Southwestern Atlantic Ocean was conducted using three distinct surface current fields~---~HYCOM, GlobCurrent, and SSalto/Duacs~---~as input to the WW3 model. 
The results underscore the sensitivity of wave fields to different current components, with significant variations in wave amplification observed across the various current inputs. 
HYCOM, incorporating ageostrophic effects and dynamic interpolation of surface currents, provided a more detailed representation of ocean surface dynamics. 
In contrast, GlobCurrent and SSalto/Duacs, derived from altimeter data, primarily captured the geostrophic component, with GlobCurrent also including the Ekman component. 
The hindcast was assessed using denoised altimeter-derived $H_s$ data at a spatial resolution of approximately 6~km~---~the high-resolution data allows for a detailed assessment of wave variability.

In the idealized simulations, a substantial increase in $H_s$ is observed in the central jet within regions of maximum opposing surface currents. 
Dipoles function as converging lenses for surface waves, effectively directing their refraction toward the central jet. 
The combined effects of wave refraction and advection significantly enhance wave energy and steepness over the central jet, with shorter period waves exhibiting more significant alterations in both height and steepness compared to longer period waves.
A substantial difference is observed in the variance distribution of $H_s$ spectra across various spatial scales when comparing conditions with and without current input. 
The $H_s$ spectrum without current input exhibits significantly lower energy~---~the variance is markedly more pronounced when currents are included into the simulations, highlighting the dynamic interplay between waves and surface currents in the vicinity of eddy dipoles.

The hindcast results indicate that the surface current from SSalto-Duacs, despite its poorer spatial and temporal resolutions, yields more reliable $H_s$ fields. 
Surface current inputs derived from ADT produce comparable effects on the wave field, with the inclusion of the Ekman component yielding no significant enhancement.
HYCOM's advantage lies in its ability to capture a wide range of dynamical processes, including ageostrophic and smaller-scale components, which are critical for accurately representing the total energy in the ocean. 
Discrepancies between the HYCOM model and the gridded SSalto/Duacs data could arise due to potential inaccuracies in the assimilation process. 
In addition to assimilation issues, every numerical model has inherent limitations, including the parameterization of sub-grid processes, errors and uncertainties in atmospheric forcing and numerical approximations~---~all of which contribute to the overall accuracy and reliability of the model output. 
While HYCOM might be less precise spatially, it provides a more comprehensive and statistically robust representation of oceanic energy dynamics. 
Although gridded altimetry might underestimate the total current components associated with eddy dipole events, it can offer a more precise representation of their position and evolution, which is particularly relevant for analyzing specific events and near real-time forecasting.

\section*{Acknowledgments}
NVC was partially supported by the Rio de Janeiro's Science Foundation (FAPERJ) under Grant Number APQ1-13/2023-293444.

\newpage

\bibliographystyle{elsarticle-harv} 
\bibliography{references_file.bib}

\end{document}